\documentclass[a4paper,fleqn,usenatbib]{mnras}

\usepackage[T1]{fontenc}
\usepackage{ae,aecompl}

\usepackage{graphicx}	
\usepackage{amsmath}	
\usepackage{amssymb}	
\usepackage{multirow}

\newcommand{\Ne}{{{n_{\rm e}}}}
\newcommand{\kB}{{{k_{\rm B}}}}
\newcommand{\id}{{{{\rm d}}}}
\newcommand{\calP}{{{\mathcal{P}}}}

\title[3-D shapes of galaxy clusters ]{The probability distribution of 3-D shapes of galaxy clusters from 2-D
X-ray images}

\author[S. Shankar \& R. Khatri]{
Swapnil Shankar,$^{1,2}$\thanks{E-mail: swapnilshankar1729@gmail.com}
Rishi Khatri$^{3}$\thanks{E-mail: khatri@theory.tifr.res.in}
\\
$^{1}$Department of Physics, UM-DAE Centre for Excellence in Basic Sciences,
Mumbai 400098, Maharashtra, India\\
$^{2}$Anton Pannekoek Institute for Astronomy, University of Amsterdam, Science Park 904, 1098 XH Amsterdam, The Netherlands\\
$^{3}$Department of Theoretical Physics, Tata Institute of Fundamental Research, Mumbai 400005, Maharashtra, India
}

\date{Accepted XXX. Received YYY; in original form ZZZ}

\pubyear{2019}

\begin{document}
\label{firstpage}
\pagerange{\pageref{firstpage}--\pageref{lastpage}}
\maketitle

\begin{abstract}
We present a new method to determine the probability distribution of the
3-D shapes of galaxy clusters from the 2-D images using stereology. In
contrast to the conventional approach of combining different data sets
(such as X-rays, Sunyaev-Zeldovich effect and lensing) to fit a 3-D model
of a galaxy cluster for each cluster, our method requires only a single
data set, such as X-ray observations or Sunyaev-Zeldovich effect
observations, consisting of sufficiently large number of clusters. Instead
of reconstructing the 3-D shape of an individual object, we recover the
probability distribution function (PDF) of the 3-D shapes of the observed
galaxy clusters. The shape PDF is the relevant statistical quantity which
can be compared with the theory and used to test the cosmological
models. We apply this method to publicly available \textit{Chandra} X-ray
data of 89 well resolved galaxy clusters. Assuming ellipsoidal shapes, we
find that our sample of  galaxy clusters is a mixture of prolate and oblate
shapes, with a preference for oblateness {with the most probable
  ratio of principle axes}  1.4 : 1.3 : 1.   The
ellipsoidal assumption is not essential to our approach and our method is
directly applicable to non-ellipsoidal shapes. Our method is insensitive to
the radial density and temperature profiles of the cluster. Our method is 
sensitive to the changes in shape of the X-ray emitting gas from inner to
outer regions and we find evidence for variation in the 3-D
shape of the X-ray emitting gas with distance from the centre.
\end{abstract}




\section{Introduction}

It has long been known that non-linear gravitational collapse in the
matter dominated Universe, starting
with Gaussian random field initial conditions, happens non-spherically, 
giving rise to Zeldovich pancakes, filaments and galaxy clusters
\citep{z1970,shz1989}, and creating the \emph{cosmic web} which has been detected in
observations \citep{gh1989,2df_survey,gott2005} as well as simulations \citep{ks1983,defw1985,millennium_simulation}.  In particular, we
expect that galaxy clusters, the largest collapsed objects in the
Universe at the intersection of filaments, will also not be perfectly
spherical \citep{fwde1988}.  Observationally we know that the galaxy
clusters are not spherical since their 2-D projections are not circular in
optical \citep{optical_non_circular_1, optical_non_circular_2}, X-ray
\citep{x_ray_non_circular_1,x_ray_non_circular_2,x_ray_non_circular_3,x_ray_non_circular_4},
Sunyaev-Zeldovich (SZ) effect \citep{sz_effect_non_circular},
weak gravitational lensing  \citep{weak_lensing_non_circular_1,weak_lensing_non_circular_2,weak_lensing_non_circular_3}
and strong gravitational lensing \citep{strong_lensing_non_circular} data. Further evidence for asphericity of
galaxy clusters comes from kinematics of galaxies in the clusters \citep{sdss_1743_clusters}. 


Cold dark matter simulations show correlations
between the
orientations of dark matter halos and the surrounding cosmic web \citep{vv1993,sm1997,numerical_triaxial_8,numerical_triaxial_7,ac2006,pc2006,am2007,brt2007}. Taking
 asphericity into account is also important for accurate mass
determinations of the galaxy clusters \citep{pj2003,cd2004,g2005,ck2007,bbp2012,sheridan2019,Huanqing2019} which in turn is important for using
the galaxy clusters for precision cosmology \citep{mantz2015,planck2016,spt2016}. The shape of
the galaxy cluster will also be influenced by the nature of dark matter,
for example the self interactions of the dark matter \citep{peter2013}. The future
X-ray \citep{erosita} and Sunyaev-Zeldovich effect surveys \citep{cmbs4} will yield hundreds of
thousands of galaxy clusters making precision cosmology with  statistics of
cluster shapes  feasible. The shape of galaxy
clusters is therefore emerging as an important observable which can be used
to test the $\Lambda$CDM cosmology, baryonic physics in the intracluster
medium (ICM) and fundamental physics such as the nature of dark matter.

One of the obstacles to using galaxy cluster shapes as a cosmological probe
is the fact that we have only 2-D information about these objects. The X-ray
and SZ effect \citep{sz_effect_classic_paper,sz1972} observations give us 2-D images in
X-ray and microwave bands. The optical galaxy surveys also do not give us
3-D information since for most galaxies we do not have absolute distance
measurements but only the redshifts. We can therefore only infer the
average distance of the cluster in a cosmological model but not the
distances to the individual galaxies. Gravitational lensing is also
mostly sensitive to the 2-D projected mass distribution. Therefore, before
we can use cluster shapes as a cosmological probe, we must solve the
problem of inferring 3-D shape of the clusters from 2-D data. 

Previously, inference of 3-D shapes of galaxy clusters, without assuming
spherical or axial symmetry, has been tried by
combining many different probes such as X-ray and SZ \citep{ls2004} with lensing data
\citep{galaxy_cluster_paper}, using
X-ray spectra \citep{samsing2012} and using weak and strong lensing data \citep{new_work_1}. We propose a new method to
infer the distribution of 3-D shapes of galaxy clusters. Our method adds a new  tool to the existing
toolkit for 3-D shape inference and can serve as an independent check of
the results obtained by other methods. Our method is relatively
computationally inexpensive and well suited to be applied to large data
sets of hundreds of thousands of clusters that will become available with
the future SZ and X-ray surveys. As we will see, our method does
not need the galaxy clusters or the gas distribution to be ellipsoidal but
can handle more general geometries. We will however make the ellipsoidal
assumption in this paper for simplicity and also to compare our results
with the published results in literature.  Similar method has been used by  \citet{main_paper} to study the neutral hydrogen gas distribution in the turbulent interstellar medium of the Milky Way. 


\section{Stereology of galaxy clusters}
{The field of Stereology combines} the ideas of Geometry and
Statistics to obtain information about the 3-D shapes of objects from a
small number of 2-D projections or cross-sections
\citep{stereology_book}. This approach lends itself naturally to
astrophysics where we usually have a single 2-D image or a
  projection of each of a large number of astrophysical objects belonging to a
  particular class or population  and we want to infer the collective 3-D properties of the
population. 
Following \citet{main_paper},  we use the probability distribution of
filamentarity  (F), a quantity constructed from Minkowski Functionals, to
solve the deprojection problem.

{In two
dimensions, the morphological properties of any 2-D
contour can be completely characterized by three Minkowski
Functionals. More
generally, in $n$ dimensions there are  $n+1$ Minkowski functionals, a
result known as Hadwidger's theorem \citep{minkowski_functional_book,minkowski_functional_cosmology}.} These three Minkowski Functionals are enclosed area $S$,
perimeter $P$ and Euler Characteristic of the contour. We will work with
simple closed contours (with Euler characteristic $=0$) so that the only
Minkowski functionals with non-trivial information are the area and the
perimeter. We can  further combine the perimeter and area to form a
quantity called Filamentarity $F$, which is defined as \citep[e.g. see][]{filamentarity_paper_example}:
\begin{equation}\label{Filamentarity}
F=\frac{P^2-4 \pi S}{P^2+4 \pi S}.
\end{equation}
This definition ensures that $0 \leq F \leq 1$ with $F=0$ corresponding to
a circle and $F\rightarrow 1$ in the limit $S\rightarrow 0$ i.e. a line
segment.  When the circle is stretched to an ellipse of increasingly higher
eccentricity, its filamentarity keeps increasing. The line segment with the
limiting value of  $F=1$ is not necessarily straight. It should be noted
that instead of filamentarity, we can also use  ellipticity as a shape
descriptor if we confine ourselves to ellipsoidal shapes. We will use
filamentarity keeping in mind future applications where we may want to
consider non-ellipsoidal shapes.

 Let us assume that a galaxy cluster is an ideal ellipsoid with length $L$,
 width $W$ and thickness $T$ ($L>W>T$). To constrain the shape, we need to
 determine the ratio $L:W:T \equiv \ell : w : 1$, where we have defined
 $\ell=L/T$ and $w=W/T$. We will work with the ratios $\ell$ and $w$, since
 we are not interested in overall size of the cluster but only its
 shape.  We
 will first consider a large sample (sample size $\sim 150,000$ clusters)
 of isotropically oriented ellipsoids (galaxy clusters) with same shape
 ($\ell$ and $w$) or equivalently observe the same galaxy cluster from
 a large number of  random observer positions. We use a simple theoretical model for X-ray emission in a
 galaxy cluster and use it to generate the theoretical X-ray surface
 brightness map for a given orientation of the cluster. We should emphasize
 that  that our
 method is not sensitive to the detailed modelling of the cluster, in particular
 the  density and temperature profiles as a function
 of distance from the centre, but only the 3-D shape of the cluster. We show this model independence
 explicitly below. We  obtain the
 isocontours of constant X-ray intensity from this X-ray surface brightness
 map and use them to  calculate the filamentarity. Finally, we repeat this
 for all the clusters of the sample to obtain the Probability Distribution
 Function (PDF) of filamentarity, $\calP(F|\ell,w)$ for a given $\ell$ and
 $w$. Comparing  the filamentarity PDF of X-ray images with the theoretical
 PDFs of different $\ell$ and $w$ then gives us information about the
 population of observed galaxy clusters. Note that by definition, $L\ge W
 \ge T$ or $\ell \ge w \ge 1$. In particular, $\ell \approx w > 1$
 corresponds to an oblate spheroid while $\ell > w \approx 1$ corresponds
 to a prolate spheroid and $\ell\approx w \approx 1$ is a spherical shape.

\subsection{The  X-ray emission model for galaxy clusters}
In galaxy clusters, with  typical temperatures $T \approx 10^7-10^8$ K, the primary emission process is thermal bremsstrahlung (free-free) emission. The total power emitted per unit volume (emissivity integrated over frequency) is given by \citep{bremsstrahlung_book}:
\begin{align}\label{emmisivity}
\epsilon = \sum_i \left(\frac{2\pi \kB}{3m}\right)^{\frac{1}{2}} \frac{2^5
  \pi e^6}{3 h m c^3} Z_i^2 \Ne n_i T^{\frac{1}{2}} \bar{g} ~ \propto
\sum_i Z_i^2 \Ne n_i T^{\frac{1}{2}}
\end{align} 
where $m$ is the mass of electron, $\Ne$ and $n_i$ are the number densities
of electrons and ions of species $i$ respectively, $Z_i$ the corresponding charge, $T$ is the electron temperature and
$\bar{g}$ is the frequency averaged Gaunt factor ($\bar{g} \sim
1.2$), $\kB$ is the Boltzmann constant, $e$ is the  charge of the electron,
$h$ is the Planck constant, and $c$ is the speed of light. Assuming charge neutrality and uniform abundance of elements
throughout the cluster gives $\sum_i Z_i^2 n_i \propto \Ne$ and therefore we
can write the X-ray surface brightness $S_X$ as integral of the emissivity
over the line of sight distance $z$, 
\begin{align}\label{X_ray_integral_initial}
S_X = \int \epsilon ~\id z  
\propto \int \Ne^2 T^{\frac{1}{2}} \id z
\end{align}

We assume a generalized  triaxial Navarro, Frenk \& White (NFW)
  model \citep{nfw_classic,numerical_triaxial_5} to represent the electron
  density $\Ne \propto \rho(R) $ of the cluster as
\begin{align}\label{X_ray_integral}
\rho(R) & = \frac{\rho_c}{\left(\frac{R}{R_S}\right)^\gamma
  \left(1+\frac{R}{R_S}\right)^{3-\gamma}},\\
R^2 & \equiv L^2
\left(\frac{x^2}{L^2}+\frac{y^2}{W^2} +\frac{z^2}{T^2}\right),\label{Eq:rdef}
\end{align}
where $x,y,z$ are the coordinates with origin at the centre of the cluster
and $R=L$ defines the outer boundary of the cluster upto which we
integrate the X-ray flux along the line of sight. Since most of the X-ray
flux is contributed by high density regions, the integral converges quickly
and we do not need to integrate out to a great distance from the cluster
centre along the line of sight. Also, we are not interested in
absolute magnitude of brightness but only the shapes of the isocontours in
an X-ray image. We have explicitly checked that the shapes of isocontours, and
hence our results, are not
sensitive to how far out we integrate as long as we integrate to a distance
greater than the scale radius $R_S$.
We  fix the values $\rho_c=1$, $\gamma=1$,
and $R_S=1$ in arbitrary units. 
The profile of electron density \citep{vkf2006} usually
  differs from the dark matter profile which the NFW model represents. Our
  observable, the filamentarity, which captures only the shape information is relatively insensitive to the exact
  density profile of gas. We check this explicitly below by using different
  electron and temperature density profiles from \citet{vkf2006} obtained
 from the X-ray   observations of different galaxy clusters. This relative insensitivity to the
  exact density profile is an advantage in our method compared to other
  methods relying on the complete 3-D modelling of the cluster.
\begin{figure}
\begin{center}
\includegraphics[width=\columnwidth]{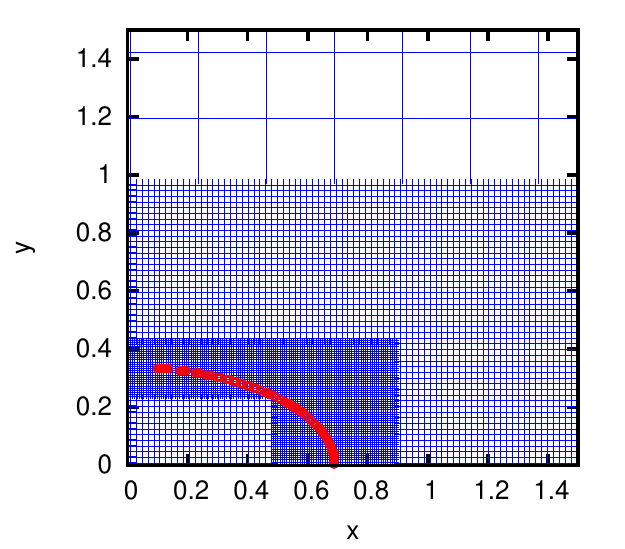}
\end{center} 
\caption{{Adaptive refinement search for the isocontours:} We start with an initial low resolution grid ($\Delta x=\Delta y = 2.26 \times 10^{-1}$ in this example). We refine the grid several times, each time increasing the resolution close to the desired isocontour, which in turn refines the isocontour. After a few refinements, we get the isocontour at high resolution ($\Delta x=\Delta y = 7.48 \times 10^{-3}$).}
\label{grid_resolution}
\end{figure}

We take a universal temperature profile for galaxy clusters \citep{temperature_profile}:
\begin{equation}\label{temp}
T(R) \propto [1+R/a_R]^{-\delta}; \qquad a_R=1, \qquad \delta = 1.6\,.
\end{equation}
To generate the X-ray surface brightness map, we start with a uniform grid in the
XY plane. We rotate the grid so that the normal to the plane of this grid
is aligned with the line of sight direction $\hat{k}$. For every point on
the rotated plane, we calculate the integral $S_X$ for X-ray intensity
using Eq.\ref{X_ray_integral_initial}. We then rotate the plane back to the
XY-plane. This gives the X-ray surface brightness map in the XY plane for
an arbitrary orientation (given by line of sight direction $\hat{k}$) of the observer with respect to the cluster. To
obtain the isocontours in a computationally efficient manner we do adaptive
refinement of the grid.
We start with an initial coarse grid  and follow the above steps
several times, each time successively increasing the resolution in the region close to
the desired isocontour. After a small number of refinements, we have the isocontour sampled at high resolution. This is illustrated in Fig.\ref{grid_resolution}. We fit the points on the isocontour  with an ellipse using
Downhill-simplex algorithm \citep{downhill_simplex}. The filamentarity of
the resultant ellipse can be easily calculated using Eq. \ref{Filamentarity}
with $S=\pi a b $ and  $P$ given with better than a percent accuracy by
\citep{l1932}
\begin{equation}\label{perimeter_formula}
 P\approx \pi(a+b)\frac{1-3\lambda^4/64}{1-\lambda^2/4},\ \ \lambda=\frac{a-b}{a+b},
\end{equation}
where $a$ and $b$ are semi-major and semi-minor axes of the ellipse respectively.

\begin{figure}
\begin{center}
\includegraphics[width=\columnwidth]{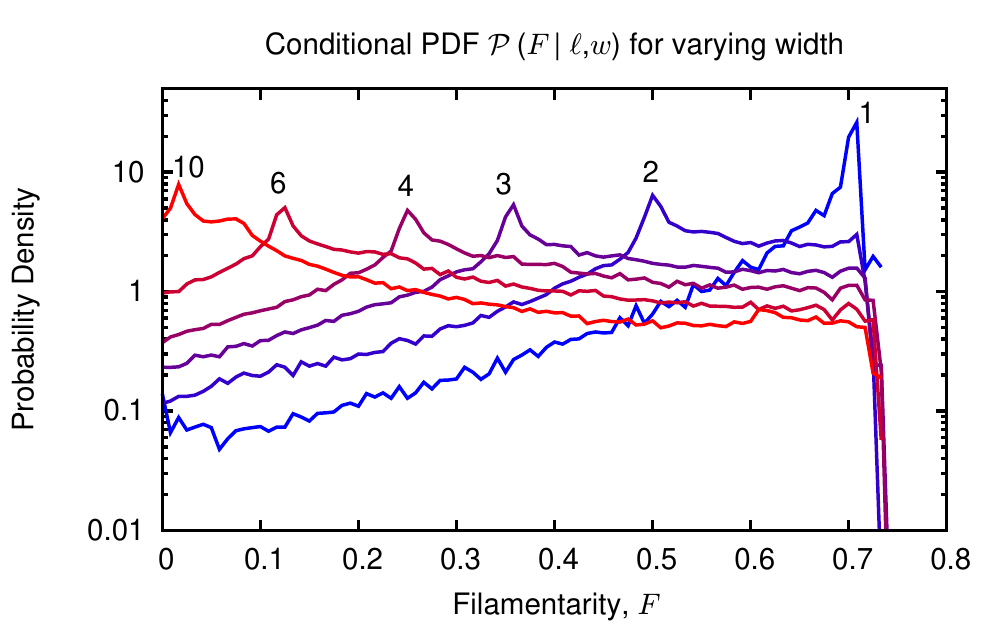}
\end{center}
\caption{Conditional filamentarity PDFs, $\calP(F|\ell,w)$, for $\ell=16$ and
  $w\in  \{1,2,3,4,6,10\}$. The PDF has
  been obtained using $\approx 150,000$ isotropic random projections binned
  into 120 intervals of $F$ between $[0,1]$. The labels refer to the values of $w$. It can be seen that peak
  filamentarity($F_p$) decreases with increase in $w$. }
\label{TPDF_varying_width}
\end{figure}

\begin{figure}
\begin{center}
\includegraphics[width=\columnwidth]{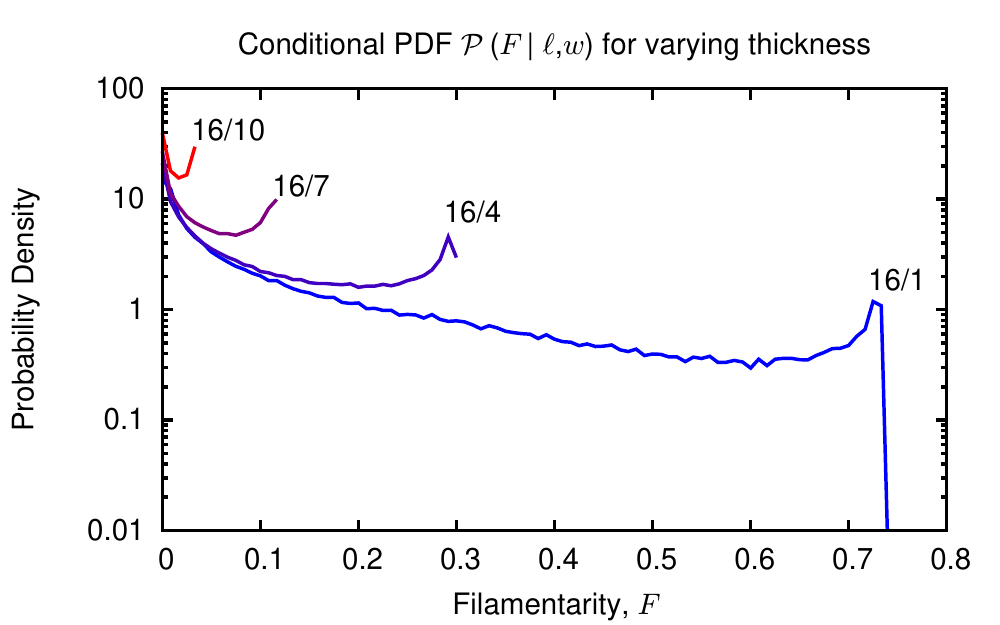}
\end{center} 
\caption{Conditional filamentarity PDFs, $\calP(F|\ell,w)$, for
  $\ell=w\in\{16,16/4,16/7,16/10\}$. The PDF has been obtained using $\approx 150,000$ isotropic random projections binned into 120 intervals of $F$ between $[0,1]$. The labels refer to the values of $\ell=w$. It can be seen that cut-off filamentarity ($F_c$) increases with increase in $\ell,w$.}
\label{TPDF_varying_thickness}
\end{figure}

\subsection{Filamentarity PDF  and its model independence}
By observing a cluster from random directions, we can build up the
probability distribution function of the filamentarity that a random
observer would measure. The filamentarity PDF for a fixed $\ell\equiv  L/T$
and $w\equiv W/T$, $\calP(F|\ell,w)$, depends on the value of $\ell,w$ of the cluster. Equivalently, if we observe single images
of large number of clusters, all of which have the same $\ell$ and $w$,
then this is the PDF we will get. The PDF  is characterized by a sharp peak
(Peak Filamentarity, $F=F_p$) and a sharp cut-off (Cut-off Filamentarity,
$F=F_c$) {which} are functions of  $\ell$ and $w$. If $L$ and $T$
are kept constant (i.e. $\ell$ constant and varying $w$), $F_p$ decreases non-linearly with an increase in
$W$. This is illustrated in Fig.~\ref{TPDF_varying_width}. On the other
hand, if $L$ and $W$ are kept constant (i.e. $\ell$ and $w$
changing by the same factor), $F_c$ decreases with an increase in
$T$ as seen in Fig.~\ref{TPDF_varying_thickness} . In reality, we expect different clusters to
have different $\ell$ and $w$ and the observed filamentarity PDF, $\calP(F)$, would be a superposition
of conditional PDFs for different $\ell$ and $w$,
\begin{align}
\calP(F) & = \int \id \ell \id w \calP(F,\ell,w)\nonumber\\
&=  \int \id \ell \id w \calP(F|\ell,w) \calP(\ell, w),\label{Eq:F}
\end{align}
where $\calP(\ell,w)$ is the PDF of shapes of clusters and $\calP(F, \ell, w)$ is
the joint  PDF.
 Our goal is to recover the PDF of shapes, 
$\calP(\ell,w)$.

\begin{figure}
\begin{center}
\includegraphics[width=\columnwidth]{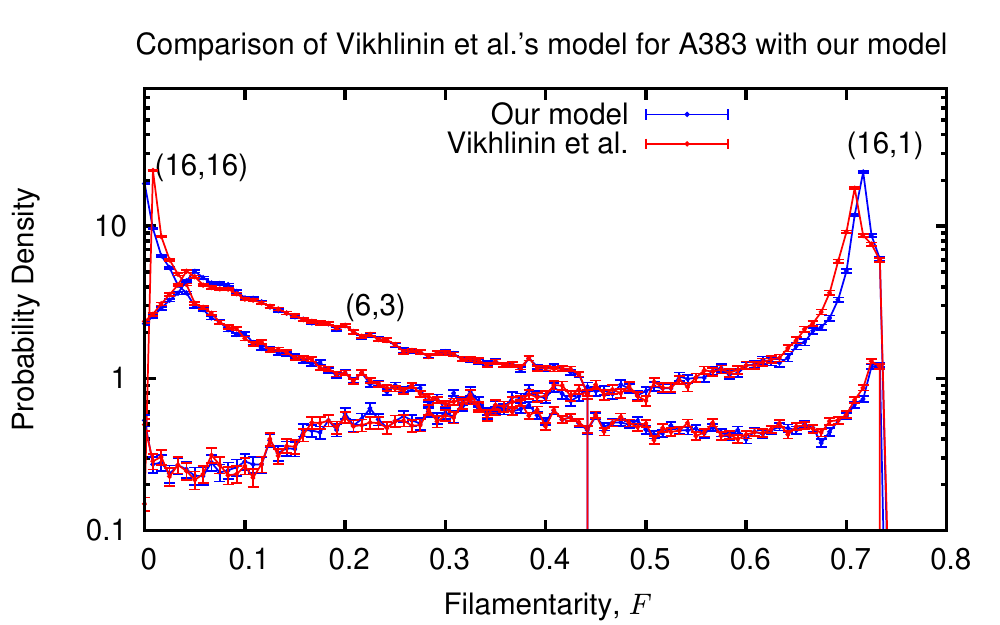}
\includegraphics[width=\columnwidth]{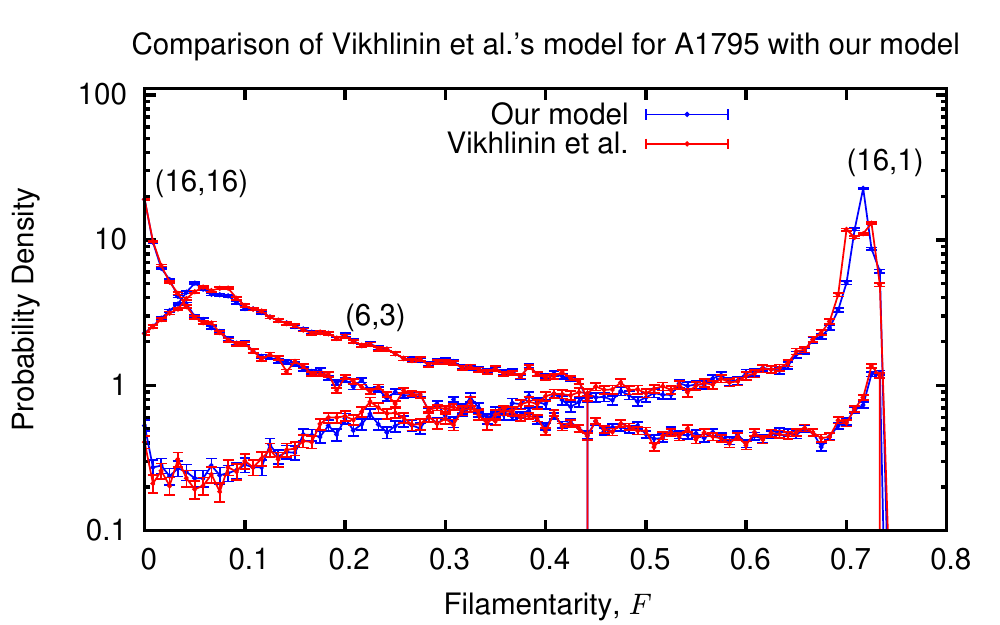}
\end{center} 
\caption{Comparison of our NFW + universal temperature profile model with temperature and density profile models from \citet{vkf2006}, for two clusters with very different parameter values: A383 and A1795. We have made the comparison for three different shapes: one prolate ($\ell=w=16$), one oblate ($\ell=16,w=1$) and one intermediate ($\ell=6,w=3$) shape. We find a general agreement of our model with that of \citet{vkf2006}, differences being less than the Monte Carlo noise.}
\label{Fig:comparison}
\end{figure}

We have used a simple model for the profiles of electron density and
temperature. Since we are interested in only the shapes of the isocontours of
X-ray surface brightness and not their overall amplitudes, our results are
not sensitive to the exact profile. To test this hypothesis, we repeat the
calculation using two different density and temperature profiles from
\citet{vkf2006} and compare it with our NFW + universal temperature profile
model in Fig. \ref{Fig:comparison}. As we can see, there is negligible
change (less than the Monte Carlo noise) in the PDF
of filamentarity confirming our assertion that the filamentarity PDF is
sensitive only to the shape of the cluster. 

\section{Filamentarity PDF of \textit{Chandra} X-ray clusters}\label{chandra_analysis}
The next step is to obtain the PDF of filamentarity of isocontours of X-ray
surface brightness maps from observational data.  We use the X-ray data of 89 galaxy clusters
 from the catalogue compiled by Eric Tittley\footnote{\url{https://www.roe.ac.uk/~ert/ChandraClusters/}} from the \textit{Chandra} Data Archive (Chaser)\footnote{\url{https://cda.harvard.edu/chaser/}} for
this purpose. {We  start by selecting an initial sample of  galaxy clusters which have
the  ratio of X-ray counts at the cluster center/peak to the background
greater than or equal to 3.} We manually
  check the cluster metadata to see if they are explicitly mentioned to be merging. We do not include these merging clusters in our sample. We use the already
processed full image data for our analysis. We do not reprocess the data
because reprocessing only changes the calibration, which will not affect
the shape of the isocontours. The list of galaxy clusters used in this
paper is given in Appendix \ref{clusterlist}. We
 should point out that we are not trying to find an \emph{average} cluster
 shape or fit a single shape to all clusters. We are inferring the full
 probability distribution function of the cluster shapes, $\calP(\ell,w)$. Therefore
 selection of clusters based on cluster physics or any other criteria a
 priori is not needed. In
 particular, for example, if there are two (or more) populations of clusters
with distinct shapes, our method would result in two (or more) distinct peaks in the
shape PDF, $\calP(\ell,w)$. {We should note that there may be large
  selection effects and we should be careful that our selection criteria
  does not remove the population that we are interested in studying. In our analysis,
we explicitly remove the merging clusters. A merging cluster can also be
roughly approximated as an elongated ellipsoid and we would expect that the shape
PDF would broaden  towards highly elongated shapes if these clusters were included. A modification
of our algorithm would be required to model merging clusters more accurately,
since they may
depart significantly from the ellipsoidal shape. Such a modification would
be non-trivial but possible since filamentarity is defined even for
non-elliptical shapes.}\\

\begin{figure}
\begin{center}
\includegraphics[scale=0.29]{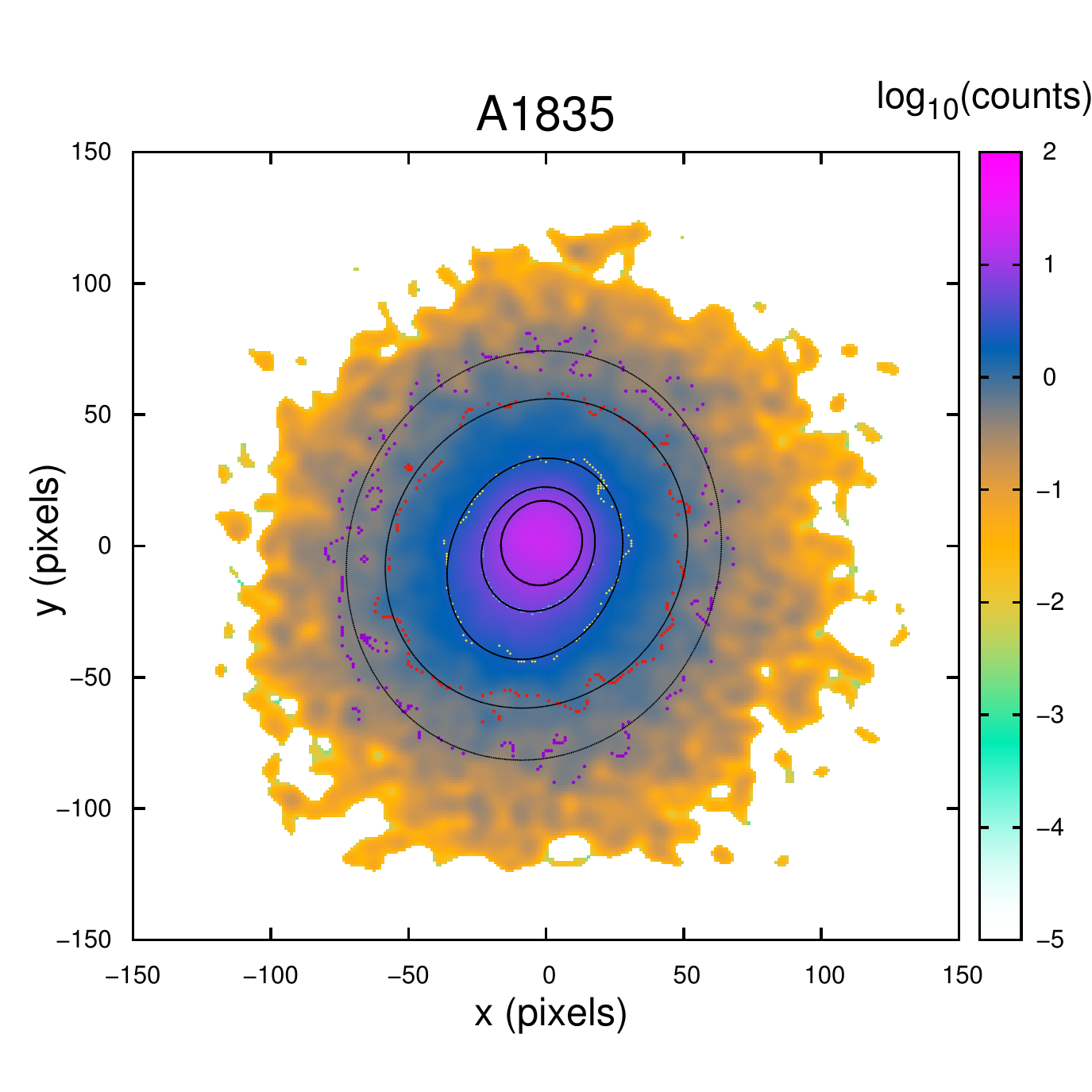}
\includegraphics[scale=0.29]{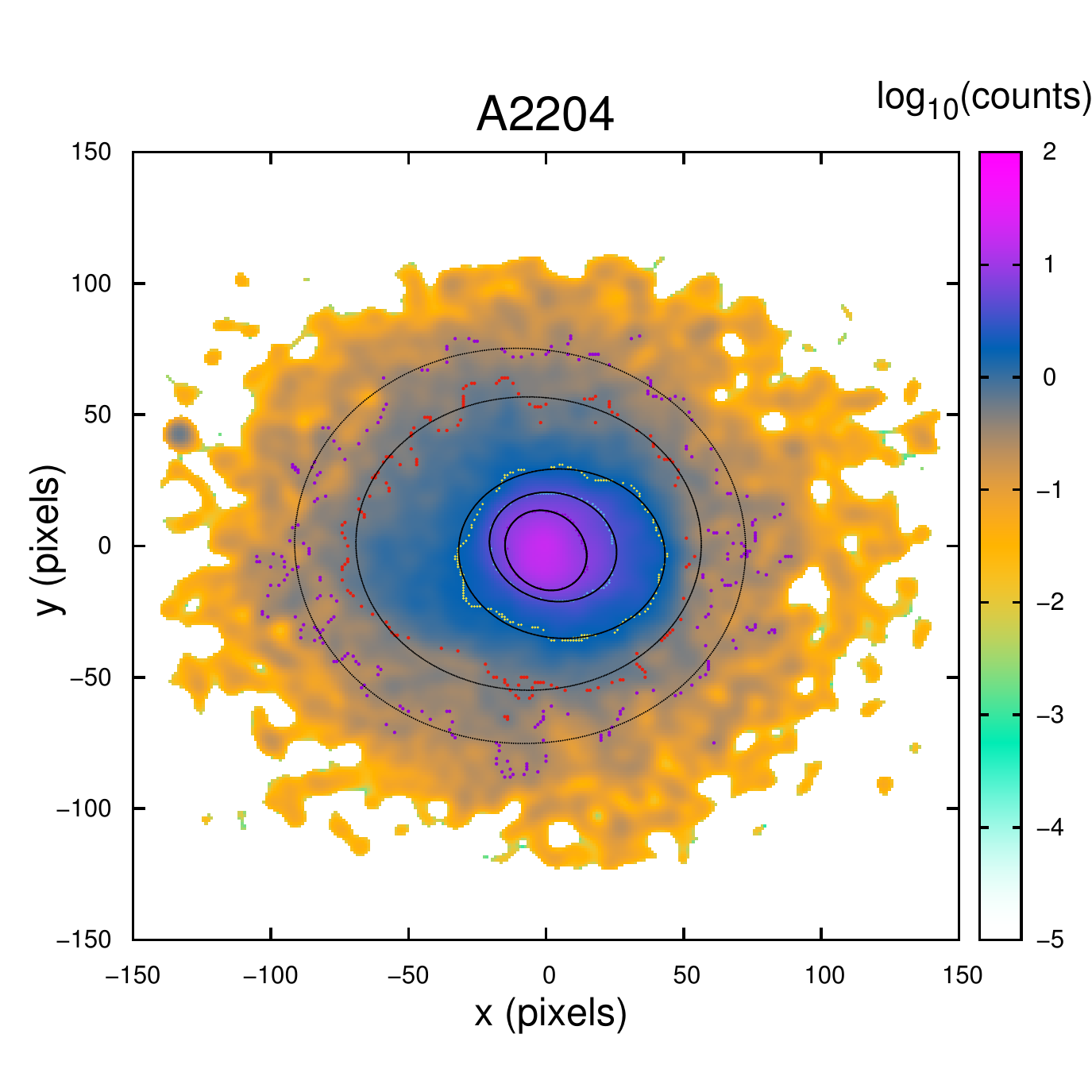}
\end{center}
\caption{{Smoothed} X-ray surface brightness maps of \textit{Chandra} clusters A1835 and A2204. The X-ray count is in log scale. The best-fitting isocontours of X-ray counts (in black color)  for 25\%,40\%,60\%,80\% and 90\% enclosed flux are shown, along with the fit points.} 
\label{isocon_pm3d}
\end{figure}

\begin{figure}
\begin{center}
\includegraphics[width=\columnwidth]{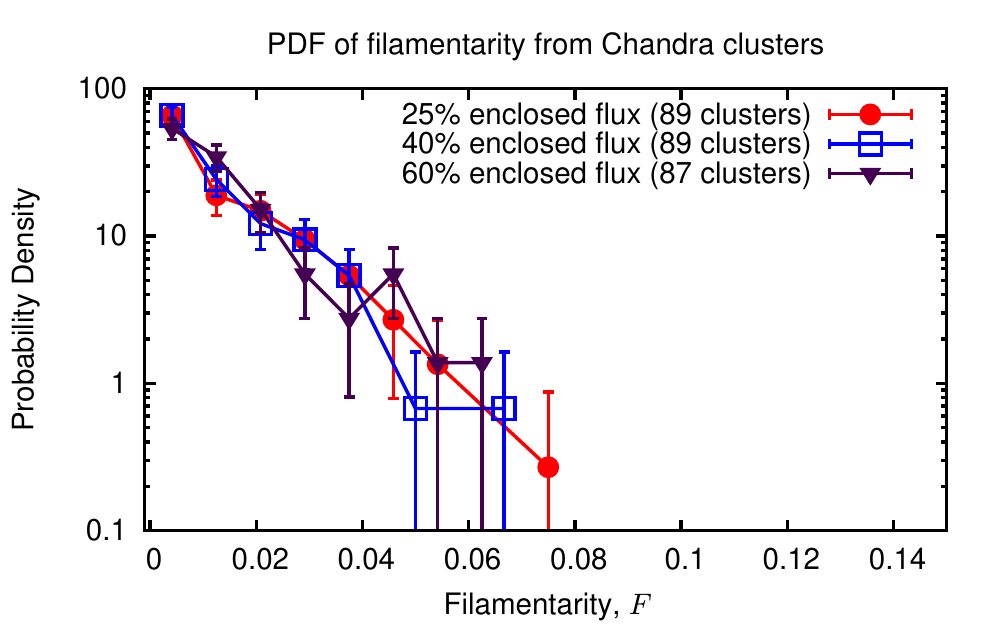}
\includegraphics[width=\columnwidth]{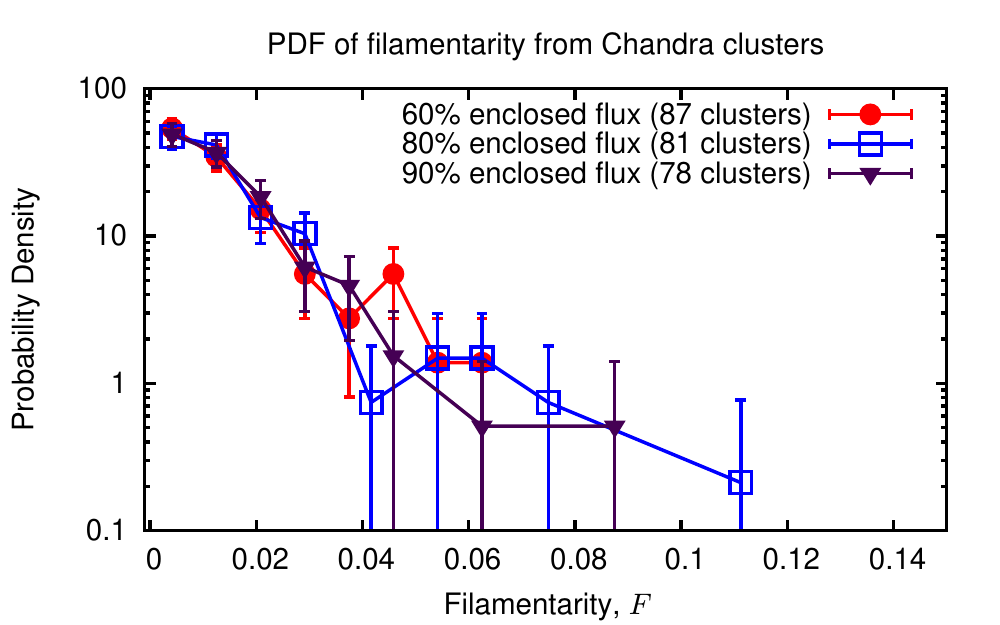}
\end{center} 
\caption{PDF of filamentarity, $\calP_{\rm obs}(F)$, from
    \textit{Chandra} X-ray clusters, obtained using up to 89
      clusters and binned with a bin-width of $1/120$ in $F$. We calculate
    the  PDF separately for the isocontours enclosing 25\%, 40\%, 60\%,
    80\% and 90\% of flux. The PDF for 80\% and 90\% flux has been obtained
    using $81$ and $78$ clusters respectively, instead of $89$, since the shapes of
    isocontours on the outer parts of a few clusters deviate
    considerably from the elliptical shape due to noise. The five
    PDF's are similar at lower values of $F$ ($F_p \approx 0$ and $F_c \approx 0.06$-$0.13$), with  differences  in the tail of the curves at larger values of $F$. }
\label{Observational_PDF}
\end{figure}

For a given cluster, we subtract the background and carry out convolution
of the resultant data using a Gaussian function with the standard deviation
$\sigma=3$
pixels {to smooth the image  and suppress noise.} After convolution, we obtain the isocontours corresponding to a
given  X-ray count and fit it with an ellipse using the Downhill-simplex algorithm. We then do a binary search by looking at isocontours corresponding to larger or smaller X-ray counts until we find the
isocontour which encloses the desired fraction of X-ray flux.  This
algorithm works because the enclosed X-ray flux  increases monotonically
and X-ray counts decrease monotonically as we move away from the
centre of the cluster. In this work, we choose isocontours which
enclose 25\%, 40\%, 60\%, 80\% and 90\% of the total flux
respectively. These isocontours laid on top of the X-ray flux maps
  of \textit{Chandra} clusters A1835 and A2204  are shown in
  Fig. \ref{isocon_pm3d}. We calculate the filamentarity of {the best-fit
ellipse using Eq. \ref{Filamentarity} and Eq. \ref{perimeter_formula}.} We repeat these steps for 89 clusters to get 89 values of
  filamentarity, which are then binned  to obtain the PDFs which are shown
in Fig.~\ref{Observational_PDF}. During this analysis, we have neglected the isocontours which are very distorted from elliptical shape. We neglect them because these isocontours are so distorted that an attempt to fit an ellipse fails for them.  This is the reason for the lower sample size when we go to isocontours of higher percent enclosed flux, because the distortion from elliptical shape is more for them. We see from Fig. \ref{Observational_PDF}
that the ellipticity of the clusters is small with the maximum filamentarity smaller than $\sim 0.15$. We have binned $F$ with a bin-width of $1/120$ and we also show the Poisson error bars. There is a difference in the tails of the PDFs, indicating that there is a variation in shape as we go from inner part of the cluster to the outer parts.

\subsection{Results for shapes of \textit{Chandra} X-ray clusters}
Our goal is to recover the shape PDF, $\calP(\ell,w)$. To this end, we can bin the shape PDF in bins of $\ell,w$ and thus reconstruct a discretized form of $\calP(\ell,w)$. This is equivalent to approximating $\calP(\ell,w)$ by a superposition of Dirac delta distributions,
\begin{align}
\calP(\ell,w) &\approx \sum_{i=1}^{n} a_i
\delta_D\left(\ell-\ell_i,w-w_i\right)
\end{align}
with the normalization condition,
\begin{align}
\sum_{i=1}^n a_i & = 1.
\end{align}
Substituting it in Eq. \ref{Eq:F}, we get after integrating out the Dirac
delta distributions,
\begin{align}
\calP(F) \approx \sum_{i=1}^{n}a_i \calP(F|\ell_i,w_i)
\end{align}

Our problem of finding the shape PDF now reduces  to finding the number of bins $n$, the bin centres $\ell_i,w_i$ and
the relative {probability} amplitudes of different shapes $a_i$ which best fit the data.

In order to proceed, we generate 100 random values of $(\ell,w)$ taken from
a uniform distribution. By definition, $\ell\ge w\ge 1$.  Hence, we take
the value of $\ell$ from a uniform random distribution $[1,2.6]$ and the
value of $w$ from a uniform random distribution $[1,\ell]$. {The
  upper limit for $\ell$, $2.6$, corresponds to $F_c \approx
  0.15$ and therefore covers the observed range of $F$.  We generate the conditional PDFs, $\calP(F|\ell,w)$, for
each of the 100 randomly sampled ($\ell,w$).} This is sufficient for the present paper since we are limited by the small size of our
data sample.

For each $w_i,\ell_i$
combination for a given $n$, we therefore find the {best-fitting} values of ($a_1,a_2,....,a_{n-1}$) by minimizing the following $\chi^2$: 
\begin{equation}\label{chisqr}
\chi^2 = \sum\limits_{j} \frac{[\calP(F_j)-\calP_{obs}(F_j)]^2}{\sigma_j^2}
\end{equation}
where $F_j$ is the $jth$ bin and $\sigma_j$ is the Poisson error in the corresponding  bin given by $\sigma_j=(\sqrt{n_{\rm obs}(j)}/n_{\rm total})\times n_{\rm bin}$, where $n_{\rm obs}$ is the observed count in that bin, $n_{\rm total}$
is the total number of clusters in the data set and $n_{\rm bin}$ is the total number of bins of filamentarity.
 
We first consider the simplest model that all clusters have the same shape,
i.e. $n=1$.  The Observational PDF has $F_p \approx 0$ and $F_c \approx
0.06-0.13$, depending on the per cent flux enclosed. We vary $w_1,\ell_1$
to find the  $\calP(F|w_1,\ell_1)$ that best fits the observed
$\calP(F)$. However, the lowest $\chi^2/d.o.f$ for this model comes out to be $\approx 1.5-3.5$, which is not satisfactory. Thus, we see immediately that the shapes of the galaxy clusters must vary from cluster to cluster. 

We next perform a  $\chi^2$-fitting of $\mathcal{P}(F)$  to the observed PDF by
fitting the $n-1$ variables $a_i$ for each combination of $w_i,\ell_i$ for
different  $n$, starting with $n=2$ and choose the combination $w_i,\ell_i$
that gives the least $\chi^2$. Note that one of the $a_i$ is fixed by the
normalization condition, $\sum_i a_i=1$. We are therefore doing a model
selection, with different  $w_i,\ell_i,n$ corresponding to a different
model for $\calP(F)$, while $a_i$ are
the parameters of the model which are being fit for each model.  
 Our model selection
consists of a brute force search for the best-fitting $w_i,\ell_i$ by repeating  the fit  for every possible set of $\ell_i,w_i$.  Thus, we have a total of $^{100}C_n$ combinations to fit. We find the $\chi^2$ for each of these sets and take the combination with the minimum value $\chi^2_{\rm min}$ as the best-fitting model. We repeat the whole procedure for $n=2,3,4$ and find $\chi^2_{\rm min}$ in each case. It should be noted that during the fitting procedure for a given $n$, we only accept those combinations of \{($\ell_i,w_i$) $i=1,n$\} which satisfy the condition that $F_c$ for at least one $\mathcal{P}(F|\ell_i,w_i)$ $\geq$ $F_c$ for $\mathcal{P}_{obs}(F)$.   The results are tabulated in Table \ref{Table_chisqr}.

To summarize, our method is accomplishing three things simultaneously:
\begin{enumerate}
\item We find the optimal number of bins, $n$, demanded by the data into which to divide the
  filamentarity PDF, $\calP(F)$.
\item For each $n$, we find the best-fitting model corresponding to different
  values of  $\{\ell_i,w_i\}$.
\item For each model we find the best-fitting parameters $a_i$ by solving the
  linear minimization problem.
\end{enumerate}

\begin{table}
 \begin{center}
\begin{tabular}{|c|c|c|c|c|c|}
\hline
 \multirow{2}{*}{$n$} & \multicolumn{5}{c|}{$\chi_{\rm min}^2/d.o.f $} \\\cline{2-6} 
        & $25\%$ flux & $40\%$ flux & $60\%$ flux & $80\%$ flux & $90\%$ flux \\ 
        \hline
1  & 3.55 & 3.99 & 1.85 & 1.19 & 1.32  \\  
2  & 1.19 & 2.61 & 0.43 & 0.80 & 0.46  \\ 
3  & 0.13 & 0.07 & 0.26 & 0.10 & 0.02   \\ 
4  & 0.01  & 0.008  &  0.19  &  0.09    & 0.005      \\ 
\hline 
\end{tabular} 
\caption{The variation of $\chi^2_{min}/d.o.f$ when Observational
    PDF ($\mathcal{P}_{obs}(F)$) at different enclosed intensities ($25\%$,
    $40\%$, $60\%$, $80\%$ and $90\%$) is fit with the theoretical  PDF
    ($\mathcal{P}(F)$) for $n=1,2,3,4$. Degree of Freedom, $d.o.f = d-p$,
    where $d$ is the number of non-zero data points and $p$ is the number
    of independent parameters. $\chi^2_{min}/d.o.f$ becomes $\lesssim 1$ at
    $n=2$ for all the cases of different enclosed X-ray flux. For 40\% flux
    case, $\chi^2_{min}/d.o.f$ is $0.07 \ll 1$ for $n=3$, which is a sign
    of overfitting, hence we choose the result for $n=2$ {for this case also.}}
 \label{Table_chisqr} 
\end{center}
 \end{table}

\begin{figure}
\begin{center}
\includegraphics[width=\columnwidth]{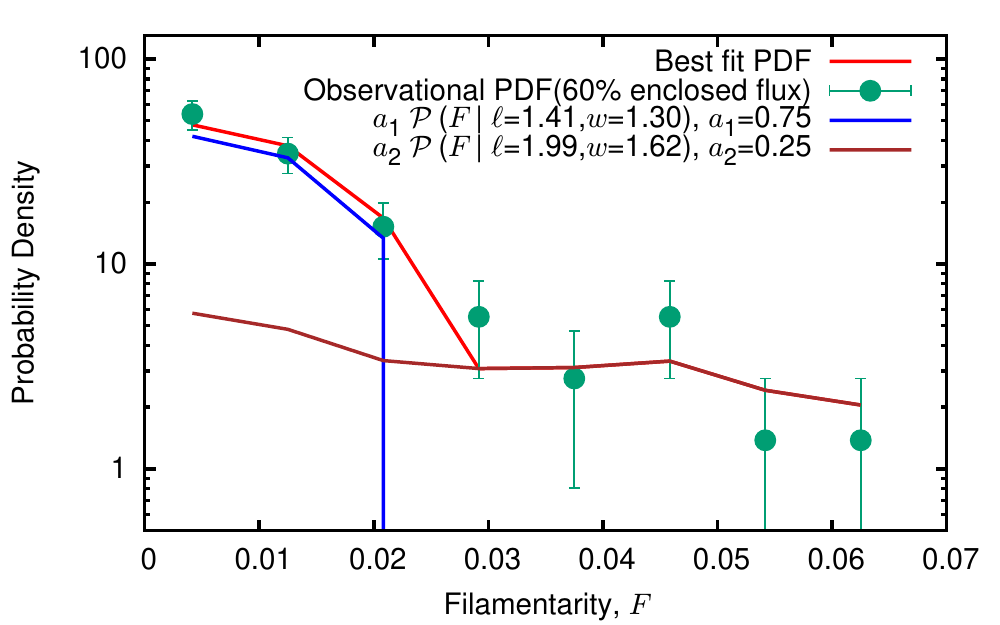}
\includegraphics[width=\columnwidth]{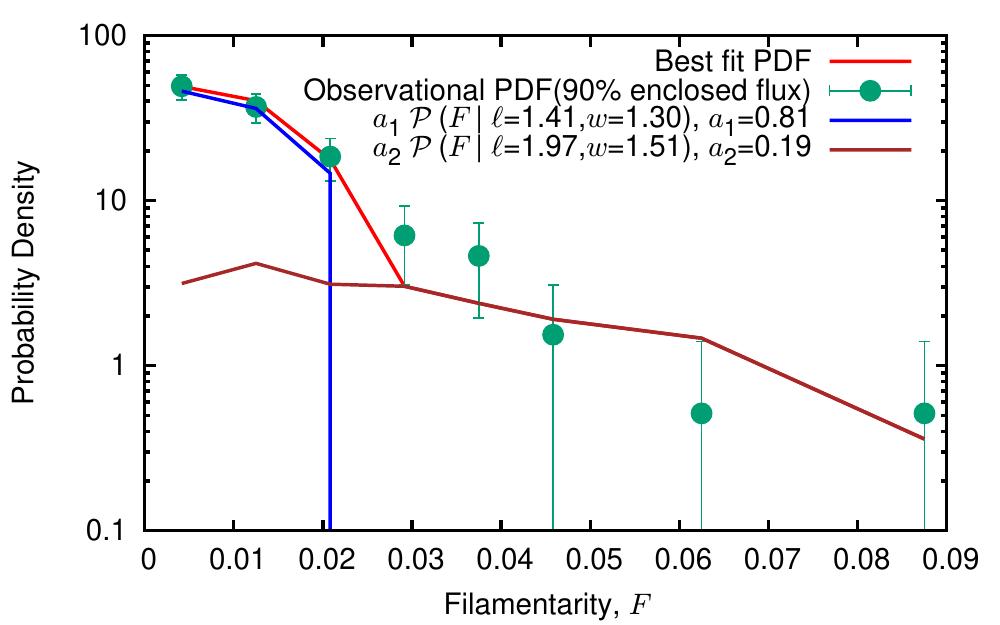}
\end{center} 
\caption{ The comparison of Observational PDF $(\mathcal{P}_{obs})$ , best fit PDF
  $\mathcal{P}(F)$ as well as the two conditional PDFs
  $\mathcal{P}(F|\ell_i,w_i)$, for $60\%$ and $90\%$ enclosed flux. We also mention the  weights $a_i$ in the legend. The shape $(\ell=1.41,w=1.30)$ contributes to the low $F$ part of both PDFs, while $(\ell=1.99,w=1.62)$ and $(\ell=1.97,w=1.51)$ contribute to the tail of the distribution for the PDF of $60\%$ and $90\%$ enclosed flux respectively.}
\label{comparison_image}
\end{figure}

We observe that $\chi^2/d.o.f$ decreases progressively from $n=1$ to $n=4$, for each enclosed per cent flux. However, $\chi^2/d.o.f$ becomes $\lesssim 1$ in every case for $n=2$. This means that $n=2$ is the optimal number of parameters. Thus, a superposition of $2$ cluster shapes is adequate to describe the data. The  {best-fitting} parameters and model for $n=2$ is shown in Table \ref{table_n=3} and Fig.\ref{comparison_image}. 

{We can}  define a triaxiality parameter $\mathcal{T}$ to classify the shapes of the clusters:
\begin{equation}
\mathcal{T} = \frac{L^2-W^2}{L^2-T^2}=\frac{\ell^2-w^2}{\ell^2-1}
\end{equation}

For a purely oblate shape, $L=W$ or $\mathcal{T}=0$. For a purely prolate
shape, $W=T$ or $\mathcal{T}=1$. We classify the shapes as nearly oblate
$(0 < \mathcal{T} \lesssim 0.33)$, triaxial $(0.33 \lesssim \mathcal{T}
\lesssim 0.67)$ or nearly prolate $(0.67 \lesssim \mathcal{T} < 1)$
\citep{numerical_triaxial_3}. The results of {this classification are} shown in Table \ref{table_n=3_shape}. We see that the higher-weighted shape is prolate for the innermost part while it has preference towards oblateness for the outer parts. Lower-weighted shape is prolate in the inner parts and triaxial in the outer parts.

\begin{table}
\begin{center}
\begin{tabular}{ccccccc}
\hline 
Enclosed flux & $\ell_1$ & $w_1$ & $a_1$ & $\ell_2$ & $w_2$ & $a_2$ \\ 
\hline 
25\%  & 1.24 & 1.07 & $0.55\pm0.13$ & 1.81 & 1.34 & 0.45 \\
 \hline 
40\%  & 1.33 & 1.24 & $0.74\pm0.12$ & 1.80 & 1.25 & 0.26  \\ 
\hline  
60\%  & 1.41 & 1.30 & $0.75\pm0.14$ & 1.99 & 1.62 & 0.25  \\ 
\hline 
80\%  & 1.41 & 1.30 & $0.84\pm0.12$ & 2.04 & 1.52 & 0.16  \\ 
\hline 
90\%  & 1.41 & 1.30 & $0.81\pm0.14$ & 1.97 & 1.51 & 0.19  \\ 
\hline 
\end{tabular} 
\caption{The best-fitting parameters and model for $n=2$ fit for $25\%$, $40\%$, $60\%$, $80\%$ and $90\%$ enclosed flux. $a_j$ represents the probability of corresponding $(\ell_j,w_j)$ in the joint PDF $\mathcal{P}(F,\ell,w)$ (Eq. \ref{Eq:F}). The error on $a_2$ is the same as that of $a_1$, since $a_2=1-a_1$. }
 \label{table_n=3}
\end{center}
\end{table}

\begin{table}
\begin{center}
\begin{tabular}{ccccc}
\hline 
Flux & $(\ell_1,w_1)$ & $\mathcal{T}_1$ & $(\ell_2,w_2)$ & $\mathcal{T}_2$ \\ 
\hline 
25\%  & (1.24,1.07) & $0.73$ (prolate) & (1.81,1.34) & $0.65$ (prolate) \\
 \hline 
40\%  & (1.33,1.24) & $0.30$ (oblate) & (1.80,1.25) & $0.75$ (prolate)  \\ 
\hline  
60\%  & (1.41,1.30) & $0.30$ (oblate) & (1.99,1.62) & $0.45$ (triaxial)  \\ 
\hline 
80\%  & (1.41,1.30) & $0.30$ (oblate) & (2.04,1.52) & $0.58$ (triaxial)  \\ 
\hline 
90\%  & (1.41,1.30) & $0.30$ (oblate) & (1.97,1.51) & $0.55$ (triaxial)  \\ 
\hline 
\end{tabular} 
\caption{Same as Table \ref{table_n=3}, but showing the shape classification of the clusters as prolate, triaxial or oblate.}
 \label{table_n=3_shape}
\end{center}
\end{table}

\subsection{Monte Carlo  estimation of the shape PDF, $\calP(\ell,w)$}\label{Monte_carlo}
 In this section we use Monte Carlo sampling to get the shape
  PDF, $\calP_{\ell,w}$.

We generate 4000 random values of $(\ell,w)$ taken from
a uniform distribution. As before, for 
 $\ell$ we have the  uniform random distribution $[1,2.6]$ and for 
 $w$ the uniform random distribution $[1,\ell]$. We generate the conditional PDFs, $\calP(F|\ell,w)$, for
each of the 4000 randomly sampled ($\ell,w$). We want to arrive at  a Monte
Carlo estimate of $\calP(\ell,w)$ with the density of points in $\ell,w$
plane $\propto \calP(\ell,w)$. 
To accomplish this, we start with the Uniform distribution. If the shape PDF was really
Uniform, we should have the filamentarity 
 distribution $\calP(F)$ given by:
\begin{equation}
\calP(F) = \sum_{i=1}^{4000}a_i \calP(F|\ell_i,w_i) \qquad \qquad a_i=1/4000
\end{equation} 
We calculate the $\chi^2$ between $\calP(F)$ and $\calP_{obs}(F)$ using
Eq. \ref{chisqr}. Next, we randomly choose 1 point out of the 4000 points
which is  removed
 and calculate the new PDF  as: 
\begin{equation}
\calP_{new}(F) = \sum_{i=1}^{3999}a_i \calP(F|\ell_i,w_i) \qquad \qquad a_i=1/3999
\end{equation} 
We calculate $\chi_{new}^2$ between $\calP_{new}(F)$ and
$\calP_{obs}(F)$. We accept or reject the change of removing the chosen
point  using the following condition :
\begin{itemize}
\item If($\chi^2_{new} \leq \chi^2 $): we accept the change and remove the
  selected point from the sample set.
\item If($\chi^2_{new} > \chi^2 $): we reject the change and do not remove
  the selected point from sample set.
\end{itemize}
We repeat this procedure for many iterations. At each iteration, we calculate the PDF as:
\begin{equation}
\calP(F) = \sum_{i=1}^{n_d}a_i \calP(F|\ell_i,w_i) \qquad \qquad a_i=1/n_d
\end{equation} 
where $n_d$ is the number of points remaining in the sample set. We stop the
procedure when $\chi^2$ converges.  Removing a point  gives higher weightage
to the remaining points. The density of points remaining in our
Monte Carlo sample set is the best estimate of the shape PDF,
$\calP(\ell,w)$.

\begin{figure}
\begin{center}
\includegraphics[scale=0.7]{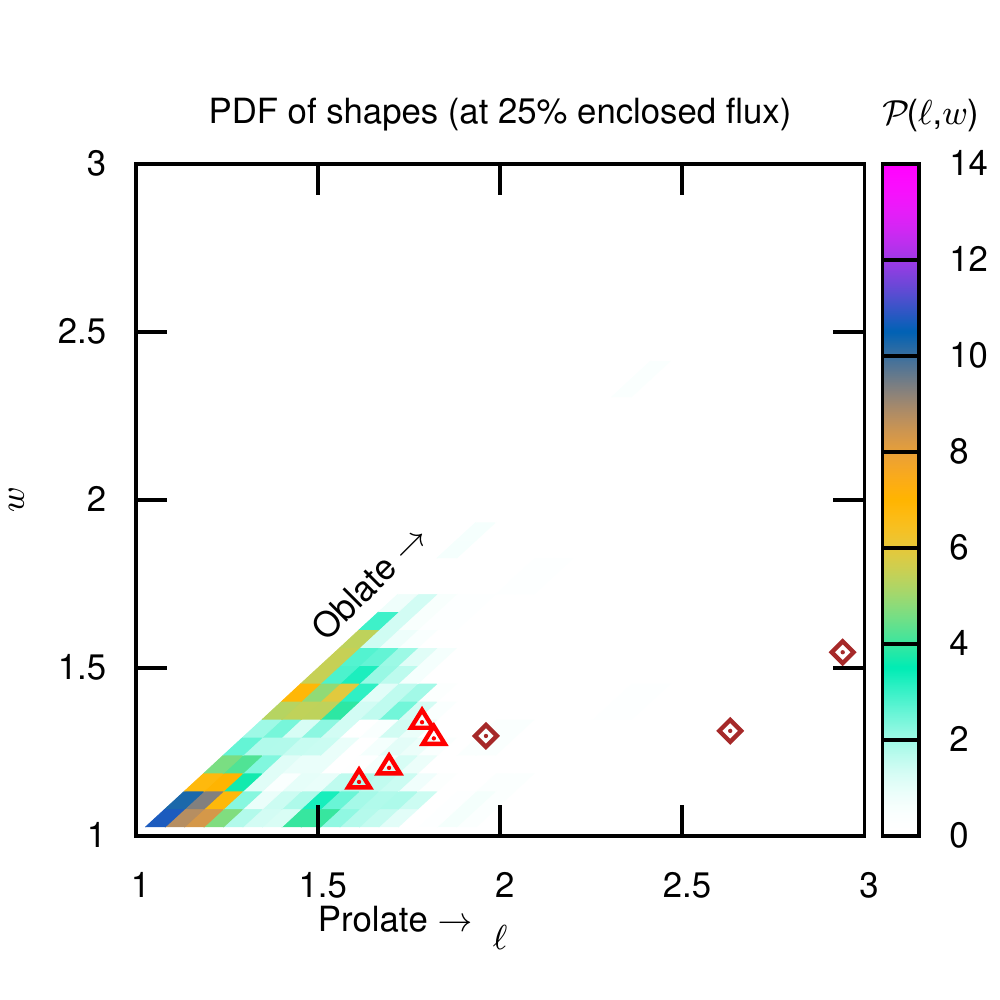}
\includegraphics[scale=0.7]{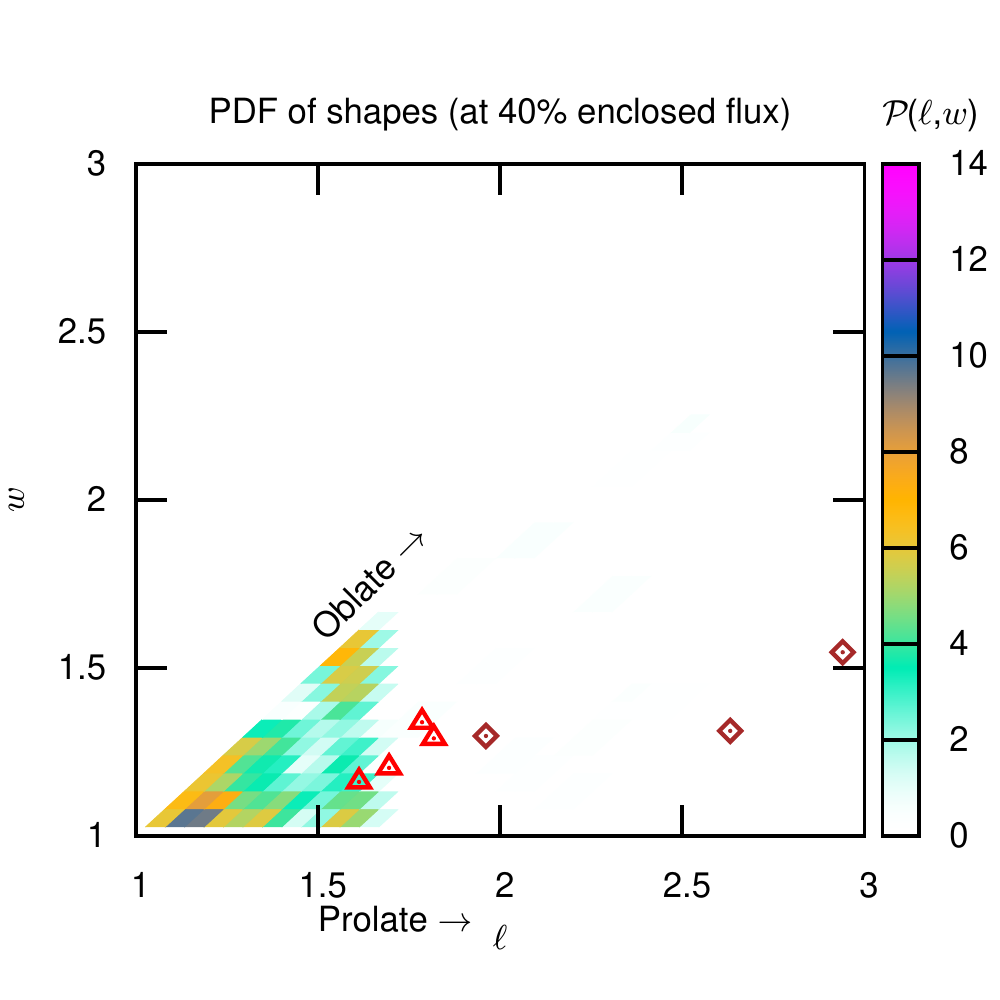}
\includegraphics[scale=0.7]{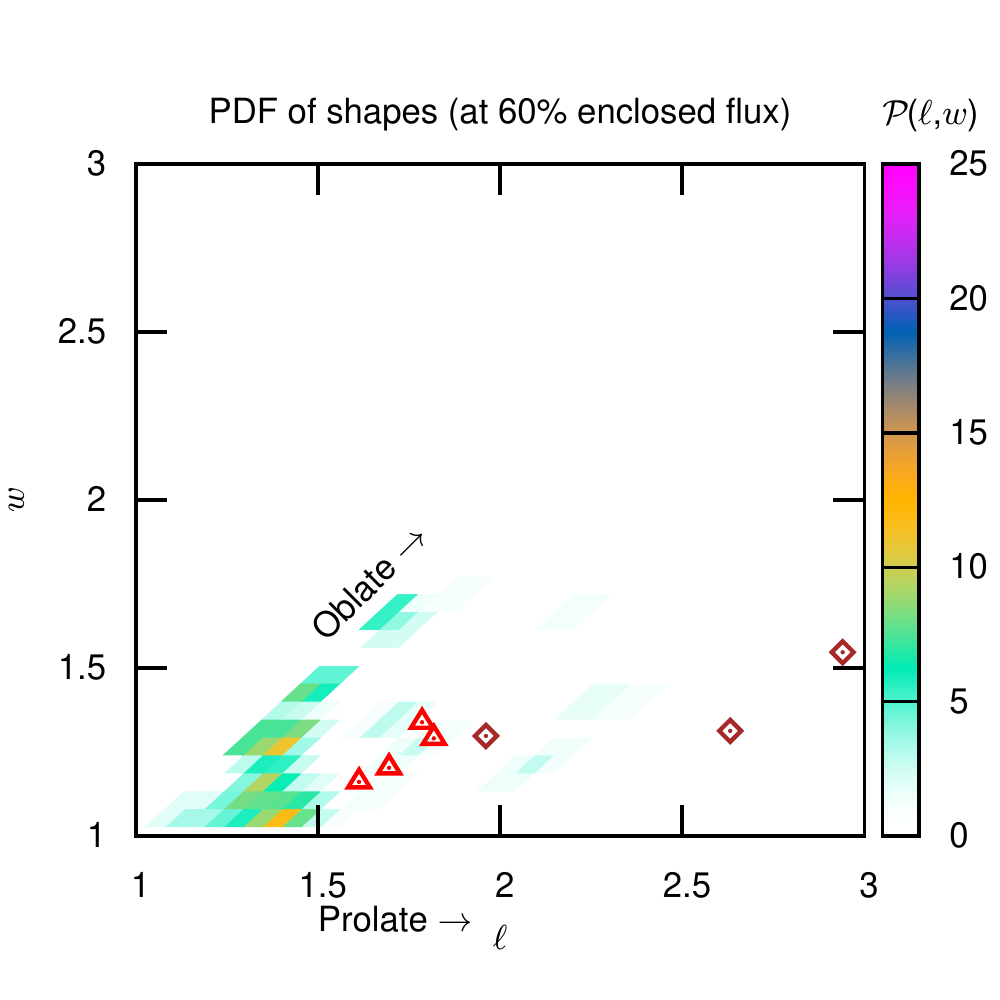}
\end{center} 
\caption{  2-D  probability density of shapes in the ($\ell,w$) plane obtained using the Monte Carlo method, for 25\%, 40\% and 60\% enclosed flux. The 4 red triangles and 3 brown diamonds correspond to the shapes of individual clusters obtained by \citet{galaxy_cluster_paper} and \citet{new_work_1} respectively. The probability density has been calculated for discrete ($\ell_j,w_j$), with a bin-width of 0.053 in $\ell,w$.   }
\label{shape_distribution}
\end{figure}

\begin{figure}
\begin{center}
\includegraphics[scale=0.7]{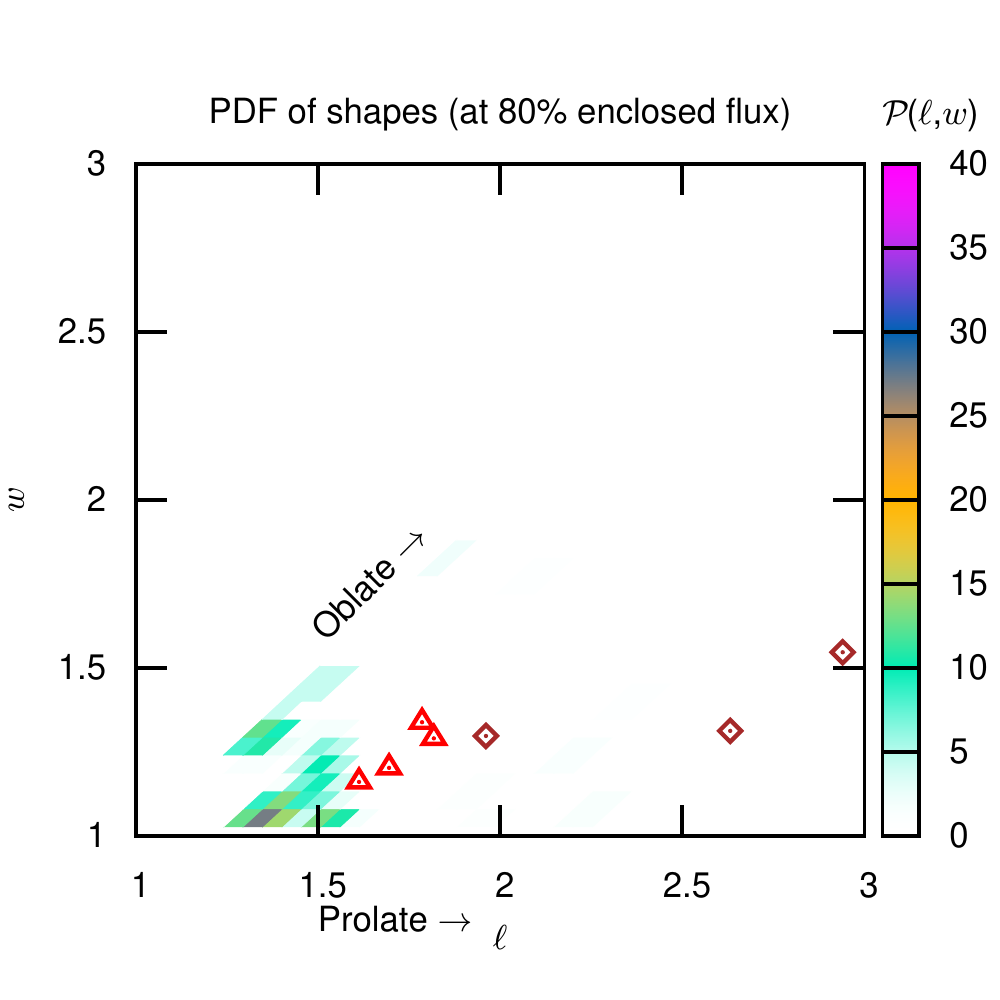}
\includegraphics[scale=0.7]{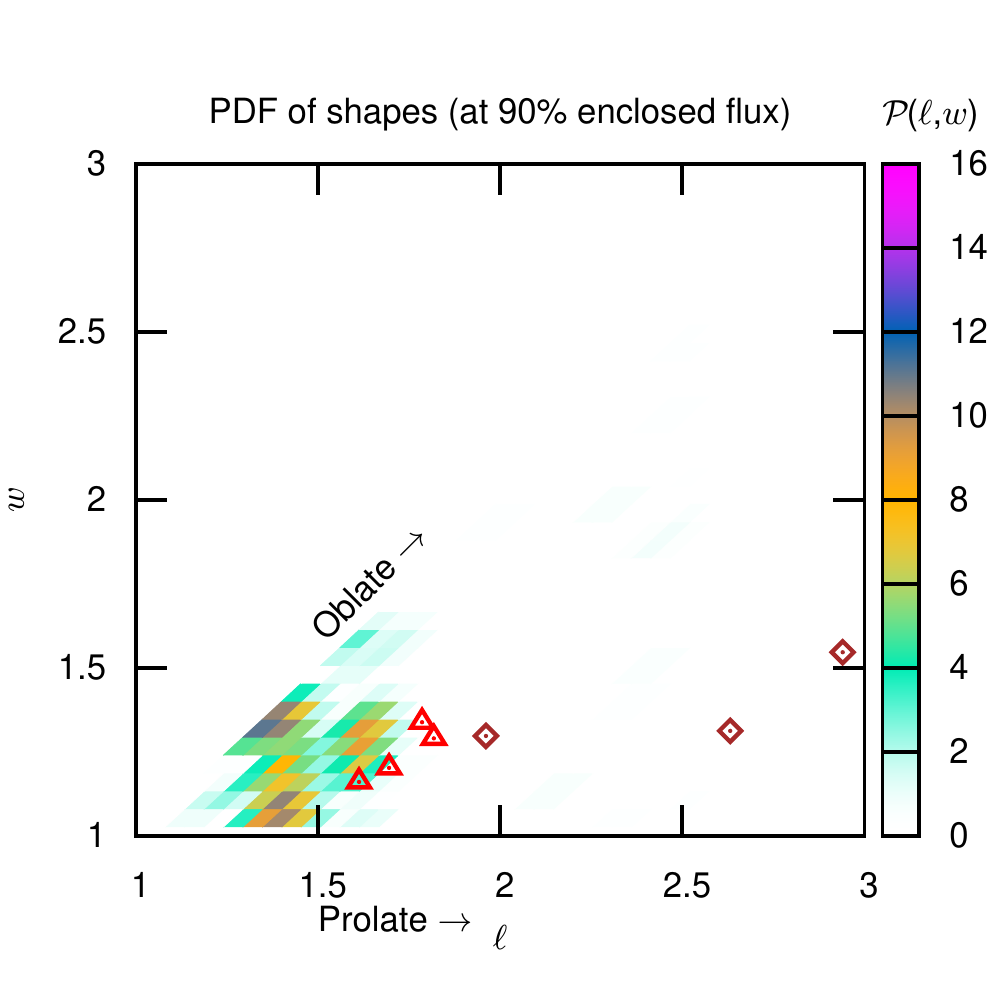}
\end{center} 
\caption{  Same as \ref{shape_distribution}, but for 80\% and 90\% enclosed flux. }
\label{shape_distribution2}
\end{figure}

We started with a uniform distribution of points in the $(\ell,w)$
  plane. After convergence, we have $\sim 100$ points non-uniformly
  distributed in the $(\ell,w)$ plane. The density of points at different
  places in the plane gives a measure of probability. It should be noted
  that in this method, the weight factor $a_i$ has been kept uniform
  throughout the calculation. Instead, the density of points per unit area
  has been used as a measure of probability. We divide the $\ell,w$ plane
  into bins of size $0.053\times 0.053$. The probability density at each
  bin center, $\ell_j,w_j$, for each bin $j$ is calculated as:
\begin{equation}
\calP(\ell_j,w_j) = k( n_j^{\rm{final}}/n_j^{\rm{initial}})/A_j
\end{equation}
where $A_j$ is a patch of square-shaped area around ($\ell_j,w_j$), $n_j^{\rm{initial}}$
is the number of points inside area $A_j$ of each square bin initially,
$n_j^{\rm{final}}$ is the number of points inside $A_j$ after convergence
at the end of Monte Carlo and $k$ is the normalization constant such that
$\sum_{j} \calP(\ell_j,w_j) A_j = 1$. The results are shown in the
form of colour maps for the 2-d PDF  in ($\ell,w$) plane in Fig. \ref{shape_distribution} and \ref{shape_distribution2}.  \\

From these figures, we see that the shapes are predominantly
  spherical in the innermost parts of a galaxy cluster, with a preference
  towards prolateness. However, as we go to the outer parts of the cluster,
  the peak shifts and the most probable shape becomes triaxial, with a
  preference towards oblateness. These results agree well with the results
  obtained in the previous section. Note that the small number of points
  remaining at convergence is just due to the fact  that our sample size is
  small.

\subsection{Comparison with previous results}
\begin{figure}
\begin{center}
\includegraphics[scale=0.68]{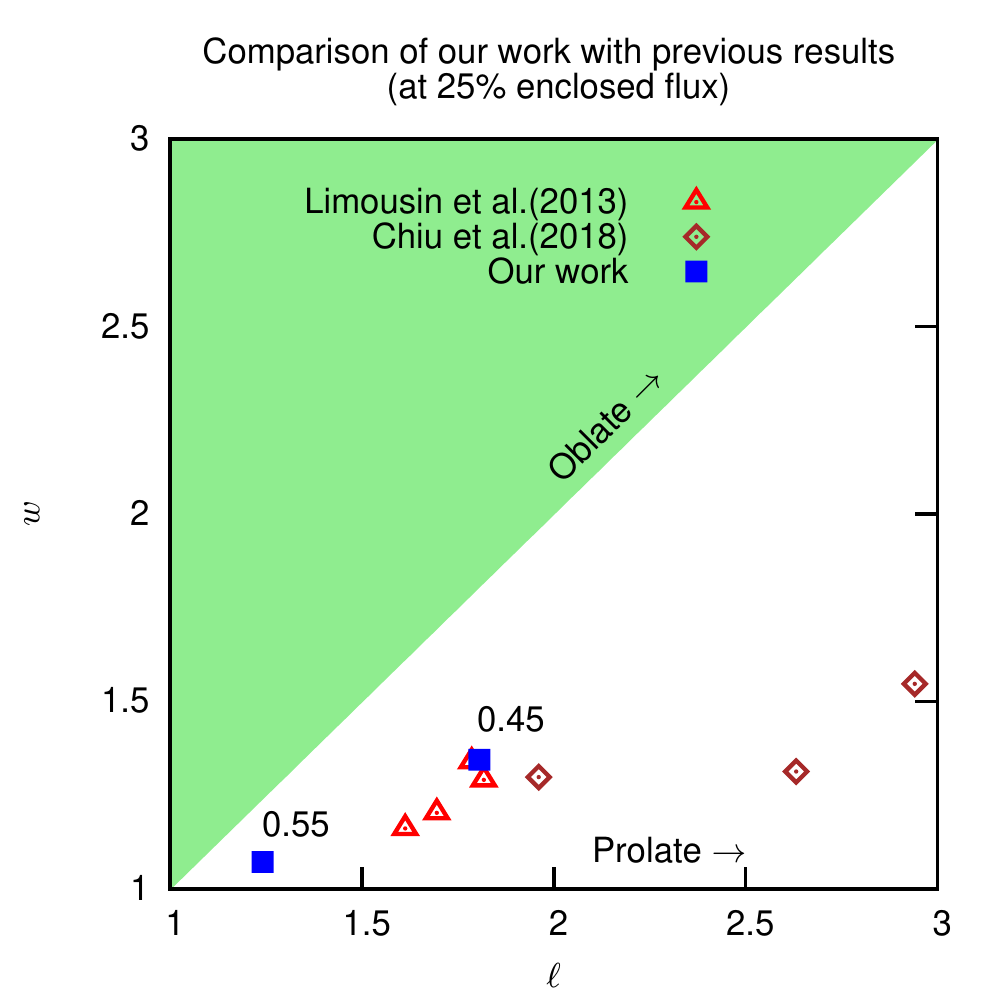}
\includegraphics[scale=0.68]{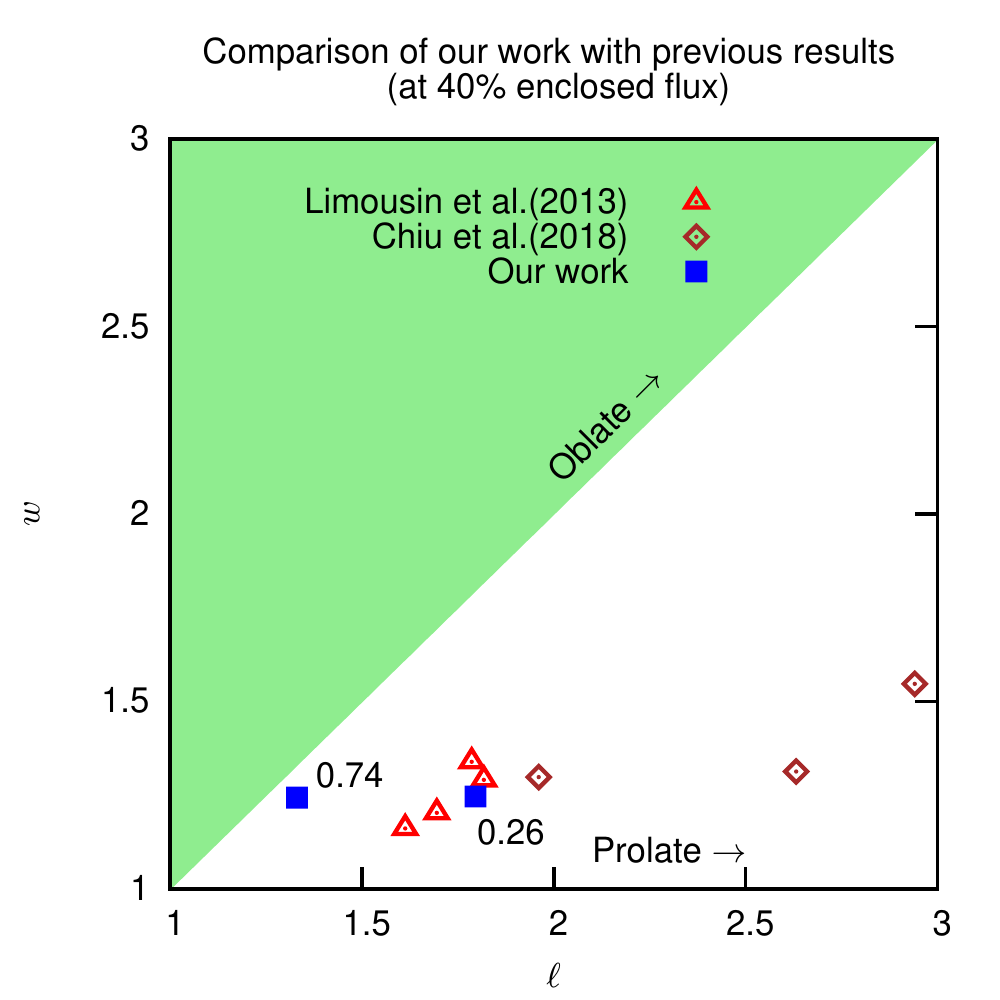}
\includegraphics[scale=0.68]{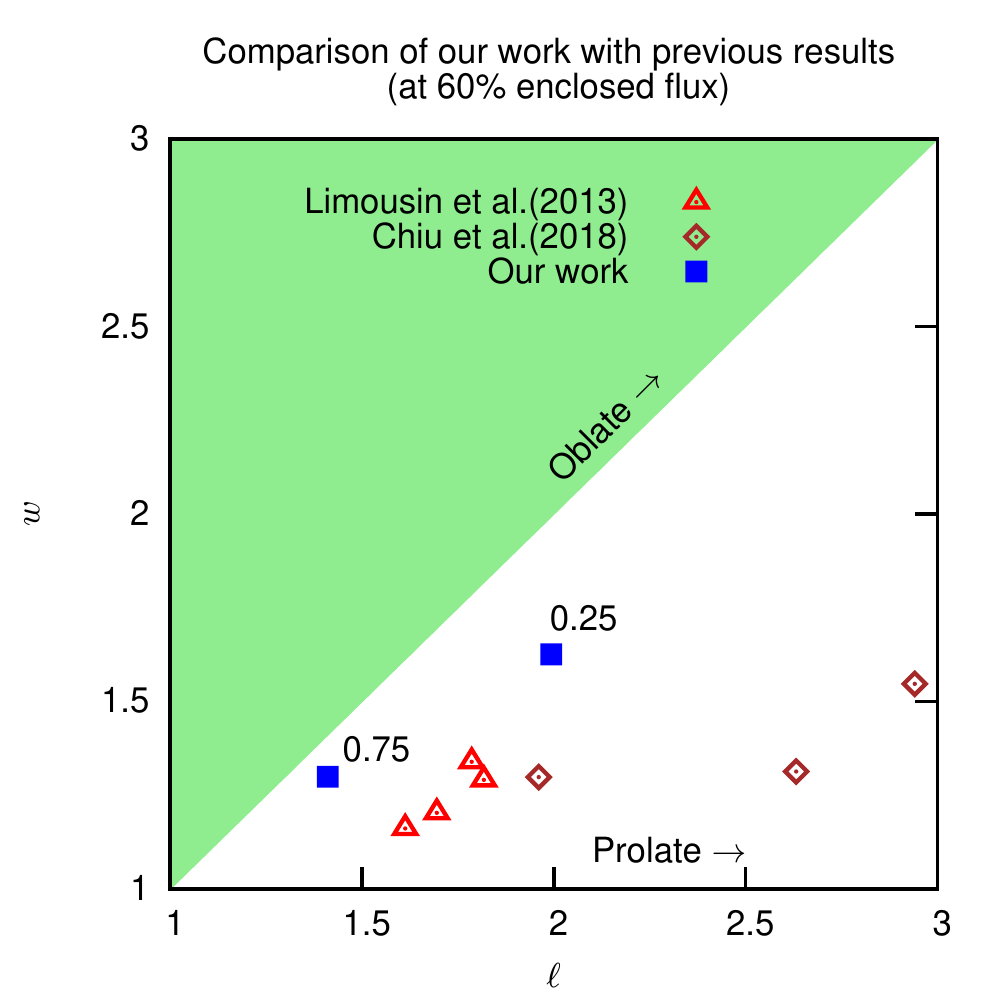}
\end{center} 
\caption{Comparison of the results obtained by us with the results of
  \citet{galaxy_cluster_paper} and \citet{new_work_1} in the $(\ell,w)$ plane. The blue square
  points obtained in this work are the approximation for the shape PDF,
  $P(\ell,w)$, of $87-89$ \emph{Chandra} clusters with the numbers next to them
  the relative {probability} amplitude of the corresponding shape. The triangles and diamonds on the other hand are the axes ratios of individual clusters from previous works. The shaded light-green region is excluded by definition of $\ell,w$. }
\label{comparison_accepted_observed}
\end{figure}

\begin{figure}
\begin{center}
\includegraphics[scale=0.68]{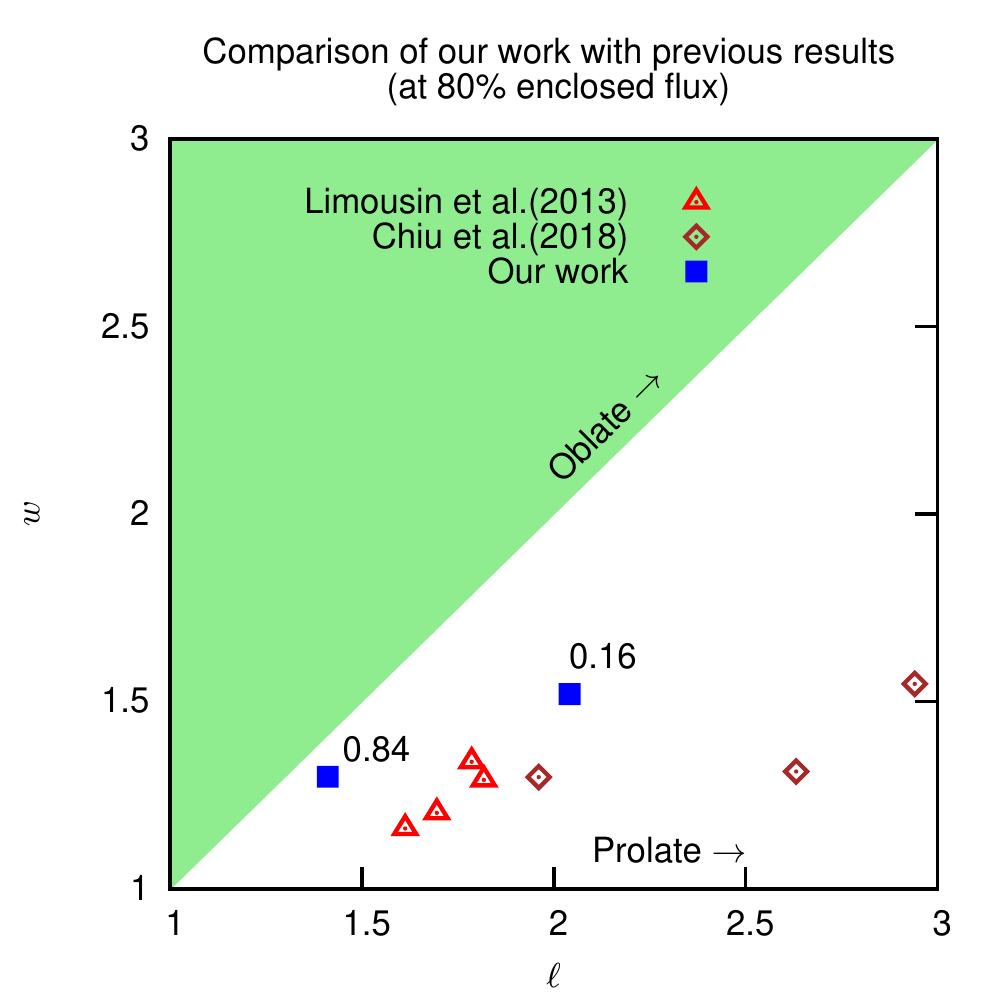}
\includegraphics[scale=0.68]{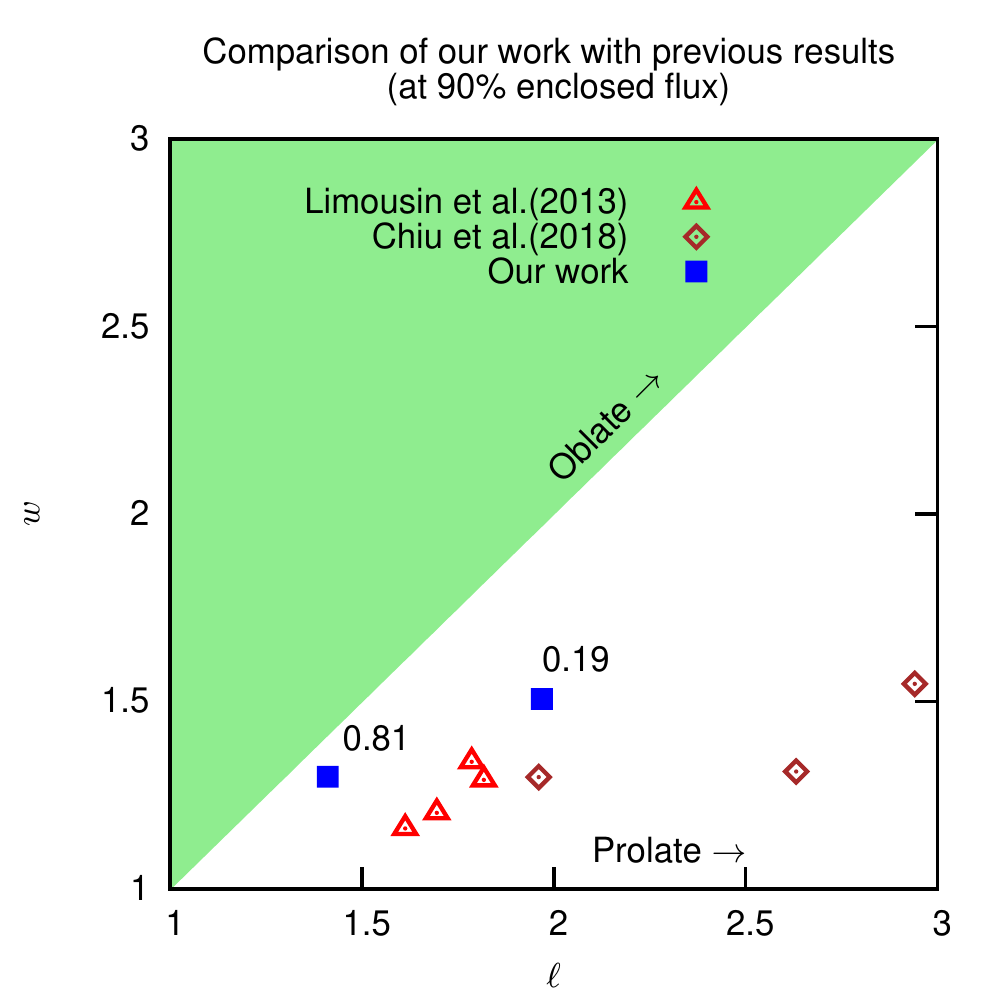}
\end{center} 
\caption{Same as \ref{comparison_accepted_observed}, but for 80\% and 90\% enclosed flux. Analysis for these two cases has been performed with $78-81$ clusters.}
\label{comparison_accepted_observed2}
\end{figure}

We compare our results with the results obtained by
\citet{galaxy_cluster_paper} and
\citet{new_work_1}. \citet{galaxy_cluster_paper} used triaxial model
fitting to obtain the 3-D shape of 4 strong lensing clusters by combining
X-ray, SZ and lensing data. \citet{new_work_1} obtained the shapes of
galaxy clusters using strong and weak lensing data. For most of the galaxy
clusters, \citet{new_work_1} are only able to constrain one of the axes
ratios and provide only lower limits on the second axes ratio when they do
not use any
shape priors from simulations.  They have given their results in the form
of $T/L$ and $W/L$, which we convert to $\ell$ and $w$ for comparison, and
tabulate the results in Table \ref{table_comparison}. We  also show the axes
ratios obtained in \citet{galaxy_cluster_paper} and
\citet{new_work_1} along with our best-fitting PDFs  in the $\ell-w$ plane in
Fig. \ref{comparison_accepted_observed} and \ref{comparison_accepted_observed2}. We should emphasize that the
points corresponding to \citet{galaxy_cluster_paper} and
\citet{new_work_1} results are the $\ell,w$ values for individual clusters
whereas the two points corresponding to our work are the two points on
the shape PDF, $P(\ell,w)$, of the 89 \emph{Chandra} clusters  with the numbers referring to the relative
{probability} amplitude of these points, $a_i$. Also our points refer to the shapes at
different distances from the {centre} for which we use the fraction of X-ray
flux enclosed by an isocontour on the X-ray surface brightness maps of the
clusters. The earlier works obtain a single average shape for the
cluster. Our results show that the shape of the halo is different as we
move from the inner regions to the outskirts of the cluster.

In the innermost part of the cluster ($25\%$ enclosed flux), we find that the cluster shapes are predominantly prolate. There is preference towards oblateness as we move to the outskirts. The points from \citet{galaxy_cluster_paper} are clustered together and are in rough agreement with our results. The points from \citet{new_work_1} are more spread out and lean towards prolate shapes. However, we note that the \citet{new_work_1} are measuring the shapes of dark matter halos while we are measuring the shape of the baryons. Taken together, these results may point towards a difference in shape of baryons and dark matter, however we need more data to make any definite conclusion.

\begin{table}
\begin{center}
\begin{tabular}{ccccc}
\hline 
Cluster & $\ell$ & $w$ & $\mathcal{T}$ & Reference \\ 
\hline 
Abell 1835  & 1.69 & 1.20 & 0.76 (prolate) & \multirow{4}{*}{\textit{L2013}} \\ 
Abell 383  & 1.82 & 1.29 & 0.71 (prolate) & \\ 
Abell 1689  & 1.79 & 1.34 & 0.64 (triaxial) & \\ 
MACS 1423  & 1.61 & 1.16 & 0.78 (prolate) & \\ 
\hline 
Abell 209 & 1.96 & 1.30 & 0.76 (prolate) & \multirow{4}{*}{\textit{C2018}} \\ 
MACS J0329-0211  & 2.94 & 1.55 & 0.82 (prolate) & \\ 
RX J1347-1145  & 2.63 & 1.31 & 0.88 (prolate) & \\ 
\hline 
\end{tabular} 
\caption{List of galaxy clusters for which shapes have been obtained by fitting cluster models to multiple data-sets (Reference \textit{L2013}: \citet{galaxy_cluster_paper},  Reference \textit{C2018}: \citet{new_work_1})}.
\label{table_comparison}
\end{center}
\end{table}

\section{Conclusions} 
We have presented a powerful new method to infer the PDF of shapes,
$P(\ell,w)$, using only 2-D images of clusters of galaxies. To illustrate
our method we have used X-ray images from publicly available \emph{Chandra}
data. Our method can also be applied to optical as well as SZ data. The
main requirement for our method is a large sample of images, since we
{directly infer the  PDF of shapes of the whole population} rather than the shapes of individual clusters. We
have also shown that our method is relatively insensitive to the density and
temperature profiles of the clusters and thus does not require detailed 3-D
modelling. 

We find general consistency with the existing results of \citet{galaxy_cluster_paper}, who use data sensitive to baryons, such as X-rays and SZ effect, given the small statistical samples but differs significantly from the analysis of \citet{new_work_1} who use only lensing data and are therefore sensitive to dark matter distribution. Our main results are that the shape of X-ray emitting gas is prolate for the innermost parts of the cluster, but shows a preference towards oblateness in the outer parts. The shapes of the haloes in dark matter simulations show preference towards prolateness in the inner parts\citep{numerical_triaxial_1,numerical_triaxial_2,
numerical_triaxial_3,numerical_triaxial_7, numerical_triaxial_11}, which is consistent with our results in the innermost parts. In these simulations, the haloes are found to be more oblate in outer parts, though prolateness is still predominant, except for \citet{numerical_triaxial_2} who find no preference towards prolate or oblate shape in the outskirts. Thus, our results in the outer parts are in contrast with the halo shapes found in dark matter simulations. However, it should again be noted that these simulation results are for the shapes of dark matter haloes, while we are measuring the shape of baryons.

We expect that the future X-ray and SZ surveys to {detect hundreds
  of thousands of galaxy clusters \citep{erosita,cmbs4}. In order to apply
  our method, we will need  these clusters to be imaged at
  high S/N by follow-up observations, similar to the Chandra cluster images
  in our selected sample.} With large statistical samples, detailed statistical comparisons of observations with simulations, including evolution of shape
with redshifts and dependence on mass, should become possible with our method. 

\section*{Acknowledgements}
We would like to thank Eugene Churazov for useful comments on the
manuscript. The scientific results reported in this article are based on data obtained
from the \textit{Chandra} Data Archive. RK would like to thank Aseem
Paranjape for interesting discussions on the subject. This work was
supported by Science and Engineering Research Board (SERB), Department of Science
and Technology, Government of India via SERB grant number
ECR/2015/000078. This work was also supported by Max-Planck-Gesellschaft
through  Max Planck Partner group between Max Planck Institute for
Astrophysics, Garching and Tata Institute of Fundamental Research, Mumbai.

\section*{Data availability}
The data underlying this article are available in Chandra Data Archive
(Chaser) at \url{https://cda.harvard.edu/chaser/}. The datasets used in
this article  were
derived from the catalogue of Chandra clusters compiled by Eric Tittley and
available at
\url{https://www.roe.ac.uk/~ert/ChandraClusters/}. The list of clusters used
in this article is given in Appendix \ref{clusterlist}.




\bibliographystyle{mnras}
\bibliography{references} 

\begin{thebibliography}{}
\makeatletter
\relax
\def\mn@urlcharsother{\let\do\@makeother \do\$\do\&\do\#\do\^\do\_\do\%\do\~}
\def\mn@doi{\begingroup\mn@urlcharsother \@ifnextchar [ {\mn@doi@}
  {\mn@doi@[]}}
\def\mn@doi@[#1]#2{\def\@tempa{#1}\ifx\@tempa\@empty \href
  {http://dx.doi.org/#2} {doi:#2}\else \href {http://dx.doi.org/#2} {#1}\fi
  \endgroup}
\def\mn@eprint#1#2{\mn@eprint@#1:#2::\@nil}
\def\mn@eprint@arXiv#1{\href {http://arxiv.org/abs/#1} {{\tt arXiv:#1}}}
\def\mn@eprint@dblp#1{\href {http://dblp.uni-trier.de/rec/bibtex/#1.xml}
  {dblp:#1}}
\def\mn@eprint@#1:#2:#3:#4\@nil{\def\@tempa {#1}\def\@tempb {#2}\def\@tempc
  {#3}\ifx \@tempc \@empty \let \@tempc \@tempb \let \@tempb \@tempa \fi \ifx
  \@tempb \@empty \def\@tempb {arXiv}\fi \@ifundefined
  {mn@eprint@\@tempb}{\@tempb:\@tempc}{\expandafter \expandafter \csname
  mn@eprint@\@tempb\endcsname \expandafter{\@tempc}}}

\bibitem[\protect\citeauthoryear{{Altay}, {Colberg}  \& {Croft}}{{Altay}
  et~al.}{2006}]{ac2006}
{Altay} G.,  {Colberg} J.~M.,   {Croft} R. A.~C.,  2006, \mn@doi [\mnras]
  {10.1111/j.1365-2966.2006.10555.x}, \href
  {https://ui.adsabs.harvard.edu/abs/2006MNRAS.370.1422A} {370, 1422}

\bibitem[\protect\citeauthoryear{{Arag{\'o}n-Calvo}, {van de Weygaert}, {Jones}
   \& {van der Hulst}}{{Arag{\'o}n-Calvo} et~al.}{2007}]{am2007}
{Arag{\'o}n-Calvo} M.~A.,  {van de Weygaert} R.,  {Jones} B. J.~T.,   {van der
  Hulst} J.~M.,  2007, \mn@doi [\apj] {10.1086/511633}, \href
  {https://ui.adsabs.harvard.edu/abs/2007ApJ...655L...5A} {655, L5}

\bibitem[\protect\citeauthoryear{Baddeley \& Jensen}{Baddeley \&
  Jensen}{2004}]{stereology_book}
Baddeley A.,  Jensen E.,  2004, Stereology for Statisticians.
Chapman \& Hall/CRC Monographs on Statistics \& Applied Probability, CRC Press,
  Boca Raton, \mn@doi{10.1201/9780203496817}

\bibitem[\protect\citeauthoryear{{Bailin} \& {Steinmetz}}{{Bailin} \&
  {Steinmetz}}{2005}]{numerical_triaxial_7}
{Bailin} J.,  {Steinmetz} M.,  2005, \mn@doi [The Astrophysical Journal]
  {10.1086/430397}, \href {http://adsabs.harvard.edu/abs/2005ApJ...627..647B}
  {627, 647}

\bibitem[\protect\citeauthoryear{{Battaglia}, {Bond}, {Pfrommer}  \&
  {Sievers}}{{Battaglia} et~al.}{2012}]{bbp2012}
{Battaglia} N.,  {Bond} J.~R.,  {Pfrommer} C.,   {Sievers} J.~L.,  2012,
  \mn@doi [\apj] {10.1088/0004-637X/758/2/74}, \href
  {https://ui.adsabs.harvard.edu/abs/2012ApJ...758...74B} {758, 74}

\bibitem[\protect\citeauthoryear{{Bett}, {Eke}, {Frenk}, {Jenkins}, {Helly}  \&
  {Navarro}}{{Bett} et~al.}{2007}]{numerical_triaxial_11}
{Bett} P.,  {Eke} V.,  {Frenk} C.~S.,  {Jenkins} A.,  {Helly} J.,   {Navarro}
  J.,  2007, \mn@doi [Monthly Notices of the Royal Astronomical Society]
  {10.1111/j.1365-2966.2007.11432.x}, \href
  {http://adsabs.harvard.edu/abs/2007MNRAS.376..215B} {376, 215}

\bibitem[\protect\citeauthoryear{{Bharadwaj}, {Sahni}, {Sathyaprakash},
  {Shandarin}  \& {Yess}}{{Bharadwaj}
  et~al.}{2000}]{filamentarity_paper_example}
{Bharadwaj} S.,  {Sahni} V.,  {Sathyaprakash} B.~S.,  {Shandarin} S.~F.,
  {Yess} C.,  2000, \mn@doi [The Astrophysical Journal] {10.1086/308163}, \href
  {http://adsabs.harvard.edu/abs/2000ApJ...528...21B} {528, 21}

\bibitem[\protect\citeauthoryear{{Binggeli}}{{Binggeli}}{1982}]{optical_non_circular_2}
{Binggeli} B.,  1982, Astronomy and Astrophysics, \href
  {http://adsabs.harvard.edu/abs/1982A\%26A...107..338B} {107, 338}

\bibitem[\protect\citeauthoryear{{Brunino}, {Trujillo}, {Pearce}  \&
  {Thomas}}{{Brunino} et~al.}{2007}]{brt2007}
{Brunino} R.,  {Trujillo} I.,  {Pearce} F.~R.,   {Thomas} P.~A.,  2007, \mn@doi
  [\mnras] {10.1111/j.1365-2966.2006.11282.x}, \href
  {https://ui.adsabs.harvard.edu/abs/2007MNRAS.375..184B} {375, 184}

\bibitem[\protect\citeauthoryear{{Buote} \& {Canizares}}{{Buote} \&
  {Canizares}}{1992}]{x_ray_non_circular_2}
{Buote} D.~A.,  {Canizares} C.~R.,  1992, in American Astronomical Society
  Meeting Abstracts \#180. p.~823

\bibitem[\protect\citeauthoryear{{Buote} \& {Canizares}}{{Buote} \&
  {Canizares}}{1996}]{x_ray_non_circular_3}
{Buote} D.~A.,  {Canizares} C.~R.,  1996, \mn@doi [Astrophysical Journal]
  {10.1086/176753}, \href {http://adsabs.harvard.edu/abs/1996ApJ...457..565B}
  {457, 565}

\bibitem[\protect\citeauthoryear{{Carter} \& {Metcalfe}}{{Carter} \&
  {Metcalfe}}{1980}]{optical_non_circular_1}
{Carter} D.,  {Metcalfe} N.,  1980, \mn@doi [Monthly Notices of the Royal
  Astronomical Society] {10.1093/mnras/191.2.325}, \href
  {http://adsabs.harvard.edu/abs/1980MNRAS.191..325C} {191, 325}

\bibitem[\protect\citeauthoryear{{Chen}, {Avestruz}, {Kravtsov}, {Lau}  \&
  {Nagai}}{{Chen} et~al.}{2019}]{Huanqing2019}
{Chen} H.,  {Avestruz} C.,  {Kravtsov} A.~V.,  {Lau} E.~T.,   {Nagai} D.,
  2019, \mn@doi [\mnras] {10.1093/mnras/stz2776}, \href
  {https://ui.adsabs.harvard.edu/abs/2019MNRAS.490.2380C} {490, 2380}

\bibitem[\protect\citeauthoryear{{Chiu}, {Umetsu}, {Sereno}, {Ettori},
  {Meneghetti}, {Merten}, {Sayers}  \& {Zitrin}}{{Chiu}
  et~al.}{2018}]{new_work_1}
{Chiu} I.-N.,  {Umetsu} K.,  {Sereno} M.,  {Ettori} S.,  {Meneghetti} M.,
  {Merten} J.,  {Sayers} J.,   {Zitrin} A.,  2018, \mn@doi [The Astrophysical
  Journal] {10.3847/1538-4357/aac4a0}, \href
  {http://adsabs.harvard.edu/abs/2018ApJ...860..126C} {860, 126}

\bibitem[\protect\citeauthoryear{{Clowe}, {De Lucia}  \& {King}}{{Clowe}
  et~al.}{2004}]{cd2004}
{Clowe} D.,  {De Lucia} G.,   {King} L.,  2004, \mn@doi [\mnras]
  {10.1111/j.1365-2966.2004.07723.x}, \href
  {https://ui.adsabs.harvard.edu/abs/2004MNRAS.350.1038C} {350, 1038}

\bibitem[\protect\citeauthoryear{{Colless} et~al.,}{{Colless}
  et~al.}{2001}]{2df_survey}
{Colless} M.,  et~al., 2001, \mn@doi [Monthly Notices of the Royal Astronomical
  Society] {10.1046/j.1365-8711.2001.04902.x}, \href
  {http://adsabs.harvard.edu/abs/2001MNRAS.328.1039C} {328, 1039}

\bibitem[\protect\citeauthoryear{{Corless} \& {King}}{{Corless} \&
  {King}}{2007}]{ck2007}
{Corless} V.~L.,  {King} L.~J.,  2007, \mn@doi [\mnras]
  {10.1111/j.1365-2966.2007.12018.x}, \href
  {https://ui.adsabs.harvard.edu/abs/2007MNRAS.380..149C} {380, 149}

\bibitem[\protect\citeauthoryear{{Davis}, {Efstathiou}, {Frenk}  \&
  {White}}{{Davis} et~al.}{1985}]{defw1985}
{Davis} M.,  {Efstathiou} G.,  {Frenk} C.~S.,   {White} S.~D.~M.,  1985,
  \mn@doi [\apj] {10.1086/163168}, \href
  {https://ui.adsabs.harvard.edu/abs/1985ApJ...292..371D} {292, 371}

\bibitem[\protect\citeauthoryear{{Dubinski} \& {Carlberg}}{{Dubinski} \&
  {Carlberg}}{1991}]{numerical_triaxial_2}
{Dubinski} J.,  {Carlberg} R.~G.,  1991, \mn@doi [Astrophysical Journal]
  {10.1086/170451}, \href {http://adsabs.harvard.edu/abs/1991ApJ...378..496D}
  {378, 496}

\bibitem[\protect\citeauthoryear{{Evans} \& {Bridle}}{{Evans} \&
  {Bridle}}{2009}]{weak_lensing_non_circular_1}
{Evans} A.~K.~D.,  {Bridle} S.,  2009, \mn@doi [The Astrophysical Journal]
  {10.1088/0004-637X/695/2/1446}, \href
  {http://adsabs.harvard.edu/abs/2009ApJ...695.1446E} {695, 1446}

\bibitem[\protect\citeauthoryear{{Fabricant}, {Rybicki}  \&
  {Gorenstein}}{{Fabricant} et~al.}{1984}]{x_ray_non_circular_1}
{Fabricant} D.,  {Rybicki} G.,   {Gorenstein} P.,  1984, \mn@doi [Astrophysical
  Journal] {10.1086/162586}, \href
  {http://adsabs.harvard.edu/abs/1984ApJ...286..186F} {286, 186}

\bibitem[\protect\citeauthoryear{{Frenk}, {White}, {Davis}  \&
  {Efstathiou}}{{Frenk} et~al.}{1988a}]{fwde1988}
{Frenk} C.~S.,  {White} S.~D.~M.,  {Davis} M.,   {Efstathiou} G.,  1988a,
  \mn@doi [\apj] {10.1086/166213}, \href
  {http://adsabs.harvard.edu/abs/1988ApJ...327..507F} {327, 507}

\bibitem[\protect\citeauthoryear{{Frenk}, {White}, {Davis}  \&
  {Efstathiou}}{{Frenk} et~al.}{1988b}]{numerical_triaxial_1}
{Frenk} C.~S.,  {White} S.~D.~M.,  {Davis} M.,   {Efstathiou} G.,  1988b,
  \mn@doi [Astrophysical Journal] {10.1086/166213}, \href
  {http://adsabs.harvard.edu/abs/1988ApJ...327..507F} {327, 507}

\bibitem[\protect\citeauthoryear{{Gavazzi}}{{Gavazzi}}{2005}]{g2005}
{Gavazzi} R.,  2005, \mn@doi [\aap] {10.1051/0004-6361:20053166}, \href
  {https://ui.adsabs.harvard.edu/abs/2005A&A...443..793G} {443, 793}

\bibitem[\protect\citeauthoryear{{Geller} \& {Huchra}}{{Geller} \&
  {Huchra}}{1989}]{gh1989}
{Geller} M.~J.,  {Huchra} J.~P.,  1989, \mn@doi [Science]
  {10.1126/science.246.4932.897}, \href
  {https://ui.adsabs.harvard.edu/abs/1989Sci...246..897G} {246, 897}

\bibitem[\protect\citeauthoryear{{Gott}, {Juri{\'c}}, {Schlegel}, {Hoyle},
  {Vogeley}, {Tegmark}, {Bahcall}  \& {Brinkmann}}{{Gott}
  et~al.}{2005}]{gott2005}
{Gott} J.~Richard I.,  {Juri{\'c}} M.,  {Schlegel} D.,  {Hoyle} F.,  {Vogeley}
  M.,  {Tegmark} M.,  {Bahcall} N.,   {Brinkmann} J.,  2005, \mn@doi [\apj]
  {10.1086/428890}, \href
  {https://ui.adsabs.harvard.edu/abs/2005ApJ...624..463G} {624, 463}

\bibitem[\protect\citeauthoryear{{Green}, {Ntampaka}, {Nagai}, {Lovisari},
  {Dolag}, {Eckert}  \& {ZuHone}}{{Green} et~al.}{2019}]{sheridan2019}
{Green} S.~B.,  {Ntampaka} M.,  {Nagai} D.,  {Lovisari} L.,  {Dolag} K.,
  {Eckert} D.,   {ZuHone} J.~A.,  2019, \mn@doi [\apj]
  {10.3847/1538-4357/ab426f}, \href
  {https://ui.adsabs.harvard.edu/abs/2019ApJ...884...33G} {884, 33}

\bibitem[\protect\citeauthoryear{Hadwiger}{Hadwiger}{1957}]{minkowski_functional_book}
Hadwiger H.,  1957, {Vorlesungen {\"U}ber Inhalt, Oberfl{\"a}che und
  Isoperimetrie}.
Berlin : Springer, \mn@doi{10.1007/978-3-642-94702-5}

\bibitem[\protect\citeauthoryear{{Jing} \& {Suto}}{{Jing} \&
  {Suto}}{2002}]{numerical_triaxial_5}
{Jing} Y.~P.,  {Suto} Y.,  2002, \mn@doi [The Astrophysical Journal]
  {10.1086/341065}, \href {http://adsabs.harvard.edu/abs/2002ApJ...574..538J}
  {574, 538}

\bibitem[\protect\citeauthoryear{{K.~N. {Abazajian} et al.}}{{K.~N. {Abazajian}
  et al.}}{2016}]{cmbs4}
{K.~N. {Abazajian} et al.} 2016, preprint, \href
  {http://adsabs.harvard.edu/abs/2016arXiv161002743A} {} (\mn@eprint {arXiv}
  {1610.02743})

\bibitem[\protect\citeauthoryear{Kasun \& Evrard}{Kasun \&
  Evrard}{2005}]{numerical_triaxial_8}
Kasun S.~F.,  Evrard A.~E.,  2005, \mn@doi [The Astrophysical Journal]
  {10.1086/430811}, 629, 781

\bibitem[\protect\citeauthoryear{{Kawahara}}{{Kawahara}}{2010}]{x_ray_non_circular_4}
{Kawahara} H.,  2010, \mn@doi [The Astrophysical Journal]
  {10.1088/0004-637X/719/2/1926}, \href
  {http://adsabs.harvard.edu/abs/2010ApJ...719.1926K} {719, 1926}

\bibitem[\protect\citeauthoryear{{Klypin} \& {Shandarin}}{{Klypin} \&
  {Shandarin}}{1983}]{ks1983}
{Klypin} A.~A.,  {Shandarin} S.~F.,  1983, \mn@doi [\mnras]
  {10.1093/mnras/204.3.891}, \href
  {https://ui.adsabs.harvard.edu/abs/1983MNRAS.204..891K} {204, 891}

\bibitem[\protect\citeauthoryear{{Lee} \& {Suto}}{{Lee} \&
  {Suto}}{2004}]{ls2004}
{Lee} J.,  {Suto} Y.,  2004, \mn@doi [\apj] {10.1086/380506}, \href
  {https://ui.adsabs.harvard.edu/abs/2004ApJ...601..599L} {601, 599}

\bibitem[\protect\citeauthoryear{{Lidstone}}{{Lidstone}}{1932}]{l1932}
{Lidstone} G.~J.,  1932, \mn@doi [{The Mathematical Gazette}]
  {10.2307/3606857}, 16

\bibitem[\protect\citeauthoryear{Limousin, Morandi, Sereno, Meneghetti, Ettori,
  Bartelmann  \& Verdugo}{Limousin et~al.}{2013}]{galaxy_cluster_paper}
Limousin M.,  Morandi A.,  Sereno M.,  Meneghetti M.,  Ettori S.,  Bartelmann
  M.,   Verdugo T.,  2013, \mn@doi [Space Science Reviews]
  {10.1007/s11214-013-9980-y}, 177, 155

\bibitem[\protect\citeauthoryear{{Loken}, {Norman}, {Nelson}, {Burns}, {Bryan}
  \& {Motl}}{{Loken} et~al.}{2002}]{temperature_profile}
{Loken} C.,  {Norman} M.~L.,  {Nelson} E.,  {Burns} J.,  {Bryan} G.~L.,
  {Motl} P.,  2002, \mn@doi [\apj] {10.1086/342825}, \href
  {https://ui.adsabs.harvard.edu/abs/2002ApJ...579..571L} {579, 571}

\bibitem[\protect\citeauthoryear{Makarenko, Fletcher  \& Shukurov}{Makarenko
  et~al.}{2015}]{main_paper}
Makarenko I.,  Fletcher A.,   Shukurov A.,  2015, \mn@doi [Monthly Notices of
  the Royal Astronomical Society: Letters] {10.1093/mnrasl/slu169}, 447, L55

\bibitem[\protect\citeauthoryear{{Mantz} et~al.,}{{Mantz}
  et~al.}{2015}]{mantz2015}
{Mantz} A.~B.,  et~al., 2015, \mn@doi [\mnras] {10.1093/mnras/stu2096}, \href
  {http://adsabs.harvard.edu/abs/2015MNRAS.446.2205M} {446, 2205}

\bibitem[\protect\citeauthoryear{{Merloni} \& {German eROSITA
  Consortium}}{{Merloni} \& {German eROSITA Consortium}}{2012}]{erosita}
{Merloni} A.,  {German eROSITA Consortium} 2012, preprint, \href
  {http://adsabs.harvard.edu/abs/2012arXiv1209.3114M} {} (\mn@eprint {arXiv}
  {1209.3114})

\bibitem[\protect\citeauthoryear{{Navarro}, {Frenk}  \& {White}}{{Navarro}
  et~al.}{1996}]{nfw_classic}
{Navarro} J.~F.,  {Frenk} C.~S.,   {White} S. D.~M.,  1996, \mn@doi [Astrophys.
  J.] {10.1086/177173}, \href
  {https://ui.adsabs.harvard.edu/\#abs/1996ApJ...462..563N} {462, 563}

\bibitem[\protect\citeauthoryear{Nelder \& Mead}{Nelder \&
  Mead}{1965}]{downhill_simplex}
Nelder J.~A.,  Mead R.,  1965, \mn@doi [The Computer Journal]
  {10.1093/comjnl/7.4.308}, 7, 308

\bibitem[\protect\citeauthoryear{{Oguri}, {Takada}, {Okabe}  \&
  {Smith}}{{Oguri} et~al.}{2010}]{weak_lensing_non_circular_2}
{Oguri} M.,  {Takada} M.,  {Okabe} N.,   {Smith} G.~P.,  2010, \mn@doi [\mnras]
  {10.1111/j.1365-2966.2010.16622.x}, \href
  {http://adsabs.harvard.edu/abs/2010MNRAS.405.2215O} {405, 2215}

\bibitem[\protect\citeauthoryear{{Oguri}, {Bayliss}, {Dahle}, {Sharon},
  {Gladders}, {Natarajan}, {Hennawi}  \& {Koester}}{{Oguri}
  et~al.}{2012}]{weak_lensing_non_circular_3}
{Oguri} M.,  {Bayliss} M.~B.,  {Dahle} H.,  {Sharon} K.,  {Gladders} M.~D.,
  {Natarajan} P.,  {Hennawi} J.~F.,   {Koester} B.~P.,  2012, \mn@doi [Monthly
  Notices of the Royal Astronomical Society]
  {10.1111/j.1365-2966.2011.20248.x}, \href
  {http://adsabs.harvard.edu/abs/2012MNRAS.420.3213O} {420, 3213}

\bibitem[\protect\citeauthoryear{{Patiri}, {Cuesta}, {Prada}, {Betancort-Rijo}
  \& {Klypin}}{{Patiri} et~al.}{2006}]{pc2006}
{Patiri} S.~G.,  {Cuesta} A.~J.,  {Prada} F.,  {Betancort-Rijo} J.,   {Klypin}
  A.,  2006, \mn@doi [\apj] {10.1086/510330}, \href
  {https://ui.adsabs.harvard.edu/abs/2006ApJ...652L..75P} {652, L75}

\bibitem[\protect\citeauthoryear{{Peter}, {Rocha}, {Bullock}  \&
  {Kaplinghat}}{{Peter} et~al.}{2013}]{peter2013}
{Peter} A. H.~G.,  {Rocha} M.,  {Bullock} J.~S.,   {Kaplinghat} M.,  2013,
  \mn@doi [\mnras] {10.1093/mnras/sts535}, \href
  {https://ui.adsabs.harvard.edu/abs/2013MNRAS.430..105P} {430, 105}

\bibitem[\protect\citeauthoryear{{Piffaretti}, {Jetzer}  \&
  {Schindler}}{{Piffaretti} et~al.}{2003}]{pj2003}
{Piffaretti} R.,  {Jetzer} P.,   {Schindler} S.,  2003, \mn@doi [\aap]
  {10.1051/0004-6361:20021648}, \href
  {https://ui.adsabs.harvard.edu/abs/2003A&A...398...41P} {398, 41}

\bibitem[\protect\citeauthoryear{{Planck Collaboration} et~al.,}{{Planck
  Collaboration} et~al.}{2016}]{planck2016}
{Planck Collaboration} et~al., 2016, \mn@doi [\aap]
  {10.1051/0004-6361/201525833}, \href
  {https://ui.adsabs.harvard.edu/abs/2016A&A...594A..24P} {594, A24}

\bibitem[\protect\citeauthoryear{{Rybicki} \& {Lightman}}{{Rybicki} \&
  {Lightman}}{1979}]{bremsstrahlung_book}
{Rybicki} G.~B.,  {Lightman} A.~P.,  1979, {Radiative processes in
  astrophysics}.
Wiley-Interscience, New York

\bibitem[\protect\citeauthoryear{{Samsing}, {Skielboe}  \& {Hansen}}{{Samsing}
  et~al.}{2012}]{samsing2012}
{Samsing} J.,  {Skielboe} A.,   {Hansen} S.~H.,  2012, \mn@doi [\apj]
  {10.1088/0004-637X/748/1/21}, \href
  {https://ui.adsabs.harvard.edu/abs/2012ApJ...748...21S} {748, 21}

\bibitem[\protect\citeauthoryear{{Sayers}, {Golwala}, {Ameglio}  \&
  {Pierpaoli}}{{Sayers} et~al.}{2011}]{sz_effect_non_circular}
{Sayers} J.,  {Golwala} S.~R.,  {Ameglio} S.,   {Pierpaoli} E.,  2011, \mn@doi
  [The Astrophysical Journal] {10.1088/0004-637X/728/1/39}, \href
  {http://adsabs.harvard.edu/abs/2011ApJ...728...39S} {728, 39}

\bibitem[\protect\citeauthoryear{{Schmalzing}, {Kerscher}  \&
  {Buchert}}{{Schmalzing} et~al.}{1996}]{minkowski_functional_cosmology}
{Schmalzing} J.,  {Kerscher} M.,   {Buchert} T.,  1996, in {Bonometto} S.,
  {Primack} J.~R.,   {Provenzale} A.,  eds, Dark Matter in the Universe,
  ProceedingsInternational School of Physics, ~Enrico Fermi~, Course 132. IOS
  press, Amsterdam, p.~281 (\mn@eprint {} {astro-ph/9508154})

\bibitem[\protect\citeauthoryear{{Shandarin} \& {Zeldovich}}{{Shandarin} \&
  {Zeldovich}}{1989}]{shz1989}
{Shandarin} S.~F.,  {Zeldovich} Y.~B.,  1989, \mn@doi [Reviews of Modern
  Physics] {10.1103/RevModPhys.61.185}, \href
  {https://ui.adsabs.harvard.edu/abs/1989RvMP...61..185S} {61, 185}

\bibitem[\protect\citeauthoryear{Skielboe, Wojtak, Pedersen, Rozo  \&
  Rykoff}{Skielboe et~al.}{2012}]{sdss_1743_clusters}
Skielboe A.,  Wojtak R.,  Pedersen K.,  Rozo E.,   Rykoff E.~S.,  2012, \mn@doi
  [The Astrophysical Journal] {10.1088/2041-8205/758/1/l16}, 758, L16

\bibitem[\protect\citeauthoryear{{Soucail}, {Fort}, {Mellier}  \&
  {Picat}}{{Soucail} et~al.}{1987}]{strong_lensing_non_circular}
{Soucail} G.,  {Fort} B.,  {Mellier} Y.,   {Picat} J.~P.,  1987, Astronomy and
  Astrophysics, \href {http://adsabs.harvard.edu/abs/1987A%26A...172L..14S}
  {172, L14}

\bibitem[\protect\citeauthoryear{{Splinter}, {Melott}, {Linn}, {Buck}  \&
  {Tinker}}{{Splinter} et~al.}{1997}]{sm1997}
{Splinter} R.~J.,  {Melott} A.~L.,  {Linn} A.~M.,  {Buck} C.,   {Tinker} J.,
  1997, \mn@doi [\apj] {10.1086/303896}, \href
  {https://ui.adsabs.harvard.edu/abs/1997ApJ...479..632S} {479, 632}

\bibitem[\protect\citeauthoryear{{Springel} et~al.,}{{Springel}
  et~al.}{2005}]{millennium_simulation}
{Springel} V.,  et~al., 2005, \mn@doi [Nature] {10.1038/nature03597}, \href
  {http://adsabs.harvard.edu/abs/2005Natur.435..629S} {435, 629}

\bibitem[\protect\citeauthoryear{{Sunyaev} \& {Zeldovich}}{{Sunyaev} \&
  {Zeldovich}}{1972}]{sz1972}
{Sunyaev} R.~A.,  {Zeldovich} Y.~B.,  1972, Comments on Astrophysics and Space
  Physics, \href {http://adsabs.harvard.edu/abs/1972CoASP...4..173S} {4, 173}

\bibitem[\protect\citeauthoryear{{Vikhlinin}, {Kravtsov}, {Forman}, {Jones},
  {Markevitch}, {Murray}  \& {Van Speybroeck}}{{Vikhlinin}
  et~al.}{2006}]{vkf2006}
{Vikhlinin} A.,  {Kravtsov} A.,  {Forman} W.,  {Jones} C.,  {Markevitch} M.,
  {Murray} S.~S.,   {Van Speybroeck} L.,  2006, \mn@doi [\apj]
  {10.1086/500288}, \href
  {https://ui.adsabs.harvard.edu/abs/2006ApJ...640..691V} {640, 691}

\bibitem[\protect\citeauthoryear{{Warren}, {Quinn}, {Salmon}  \&
  {Zurek}}{{Warren} et~al.}{1992}]{numerical_triaxial_3}
{Warren} M.~S.,  {Quinn} P.~J.,  {Salmon} J.~K.,   {Zurek} W.~H.,  1992,
  \mn@doi [Astrophysical Journal] {10.1086/171937}, \href
  {http://adsabs.harvard.edu/abs/1992ApJ...399..405W} {399, 405}

\bibitem[\protect\citeauthoryear{{Zeldovich}}{{Zeldovich}}{1970}]{z1970}
{Zeldovich} Y.~B.,  1970, \aap, \href
  {http://adsabs.harvard.edu/abs/1970A%26A.....5...84Z} {5, 84}

\bibitem[\protect\citeauthoryear{{Zeldovich} \& {Sunyaev}}{{Zeldovich} \&
  {Sunyaev}}{1969}]{sz_effect_classic_paper}
{Zeldovich} Y.~B.,  {Sunyaev} R.~A.,  1969, \mn@doi [Astrophysics and Space
  Science] {10.1007/BF00661821}, \href
  {http://adsabs.harvard.edu/abs/1969Ap\%26SS...4..301Z} {4, 301}

\bibitem[\protect\citeauthoryear{{de Haan} et~al.,}{{de Haan}
  et~al.}{2016}]{spt2016}
{de Haan} T.,  et~al., 2016, \mn@doi [\apj] {10.3847/0004-637X/832/1/95}, \href
  {http://adsabs.harvard.edu/abs/2016ApJ...832...95D} {832, 95}

\bibitem[\protect\citeauthoryear{{van Haarlem} \& {van de Weygaert}}{{van
  Haarlem} \& {van de Weygaert}}{1993}]{vv1993}
{van Haarlem} M.,  {van de Weygaert} R.,  1993, \mn@doi [\apj]
  {10.1086/173416}, \href
  {https://ui.adsabs.harvard.edu/abs/1993ApJ...418..544V} {418, 544}

\makeatother
\end{thebibliography}




\appendix
\section{Error propagation from Chandra X-ray Data and robustness of
  filamentarity PDFs}

We can test the robustness of our method to noise in X-ray data as well as
intrinsic departures of the cluster shapes from perfect ellipsoids by
propagating the errors in the ellipse parameters from the X-ray images to
the filamentarity PDFs. The noise in X-ray data as well as non-linear
cluster physics would result in making the isocontours irregular. The
scatter of the isocontours around the best-fit ellipse and  the resulting errors
on the ellipse filamentarity, therefore, capture both the important sources of
error. We will call the departure of an isocontour from a perfect ellipse (or
scatter of the isocontour points around the best fit ellipse) as
\emph{isocontour noise}.  We show below explicitly that, for our sample, these errors are negligible
compared to the Poisson errors due to the limited sample size. This
analysis justifies
our approach of ignoring these errors in the main text. In future, as more
and more clusters are imaged in X-rays, the Poisson errors may become
comparable with the isocontour noise with increasing sample size, and it will be important to propagate
these errors to final results. As we show below, this is quite
straightforward to implement. Furthermore, we can use the errors in
filamentarity, $F$, as an additional selection criteria to remove the most
irregular clusters before performing the shape analysis. 

For each cluster listed in Appendix \ref{clusterlist}, we obtain the isocontour corresponding to a given X-ray count. We fit this isocontour with an ellipse as described in detail in section \ref{chandra_analysis}.  The fitting procedure gives us the best-fitting values of semi-major axis, $a$, and semi-minor axis, $b$, of the resulting ellipse, along with the covariance matrix, $cov(a,b)$, associated with the fit. We find that the error on $a$ and $b$ is small, typically $1$-$2\%$ in most cases. In this section, we propagate the error on $a$ and $b$ to the PDF of filamentarity of Chandra clusters, $\calP_{\rm obs}(F)$.

\begin{figure}
\begin{center}
\includegraphics[width=\columnwidth]{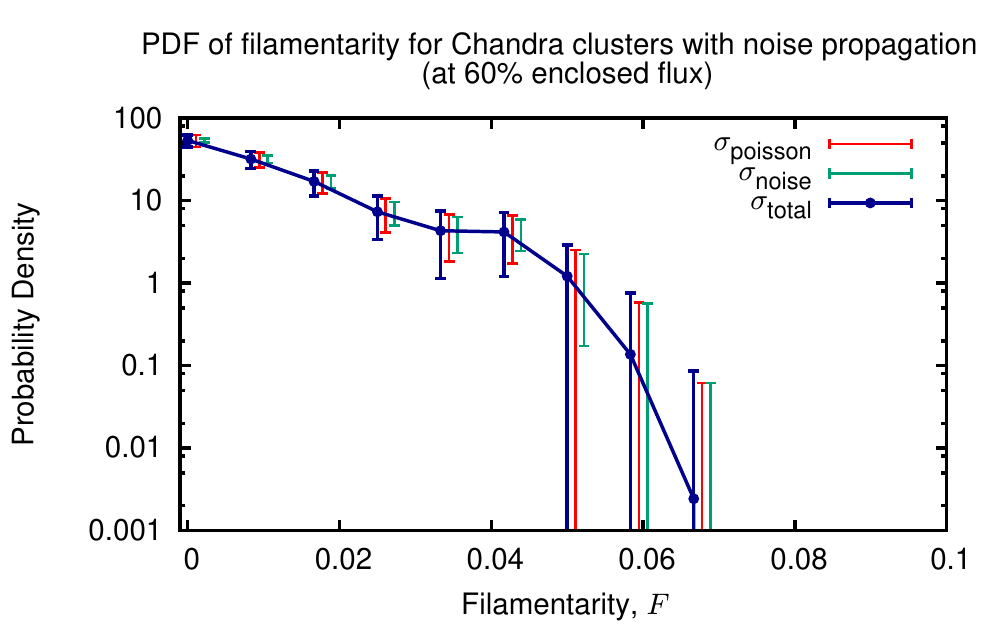}
\end{center} 
\caption{  PDF of filamentarity, $\calP_{\rm obs}(F)$, for
    \textit{Chandra} X-ray clusters for 60\% enclosed flux, obtained using 83
      clusters after propagation of errors from isocontour noise  and
      binned with a bin-width of $1/120$ in $F$. The rejection criterion
      leads to the rejection of 4 clusters from the original sample for
      which error in filamentarity ($\sigma_F$) is greater than the bin width. The 2 components of error, $\sigma_{poisson}$ and $\sigma_{noise}$, are shown alongside the total error, $\sigma_{total}$. Poisson error, which comes from the finite number of clusters, has the dominant contribution in the total error. }
\label{error_propagated_PDF60}
\end{figure}
 
For a given cluster, we choose a random value $(a',b')$ from a 2D joint
Gaussian probability distribution with mean ($a,b$) and covariance matrix
$cov(a,b)$. We use $(a',b')$ to calculate the random sample  of
filamentarity $F'$ for the corresponding cluster.  We repeat this for every
cluster and obtain the random sample of PDF of filamentarity, $\calP_{\rm
  obs}(F')$.  We  repeat the above steps 10000 times to generate 10000
samples  $\calP_{\rm obs}(F')$ from the X-ray images of our cluster
sample. In order to calculate the final PDF $\calP_{\rm obs}(F)$, we take
the mean and standard deviation of probability density values in each
filamentarity bin across the 10000 PDFs. For a given bin, mean gives the
probability density and standard deviation gives the propagated error from
the isocontour noise. For each cluster, the standard deviation of the samples
$F'$ from the mean  gives the error on
 the filamentarity of the corresponding cluster, $\sigma_{F}$. We can use
a threshold on $\sigma_F$ to reject clusters with very irregular
isocontours. For example, we would expect  clusters which are not in virial
equilibrium or are merging to have significant irregularities and departures from elliptical
shapes.  For the analysis in this appendix, we use the selection criteria
that the error on the filamentarity should be less than our chosen bin width for $F$
and thus reject clusters with $\sigma_F>1/120$.

\begin{figure}
\begin{center}
\includegraphics[scale=0.3]{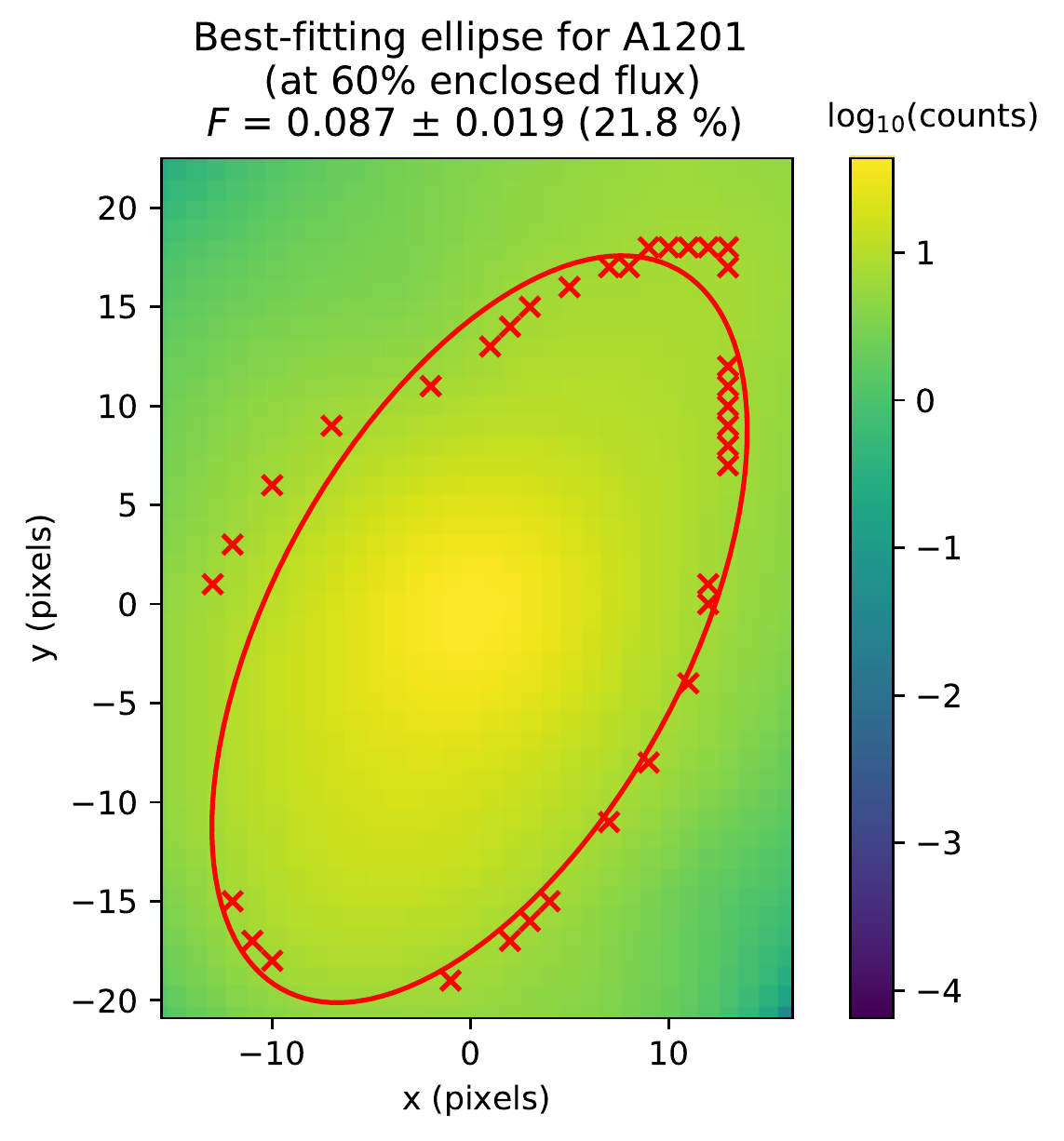}
\includegraphics[scale=0.3]{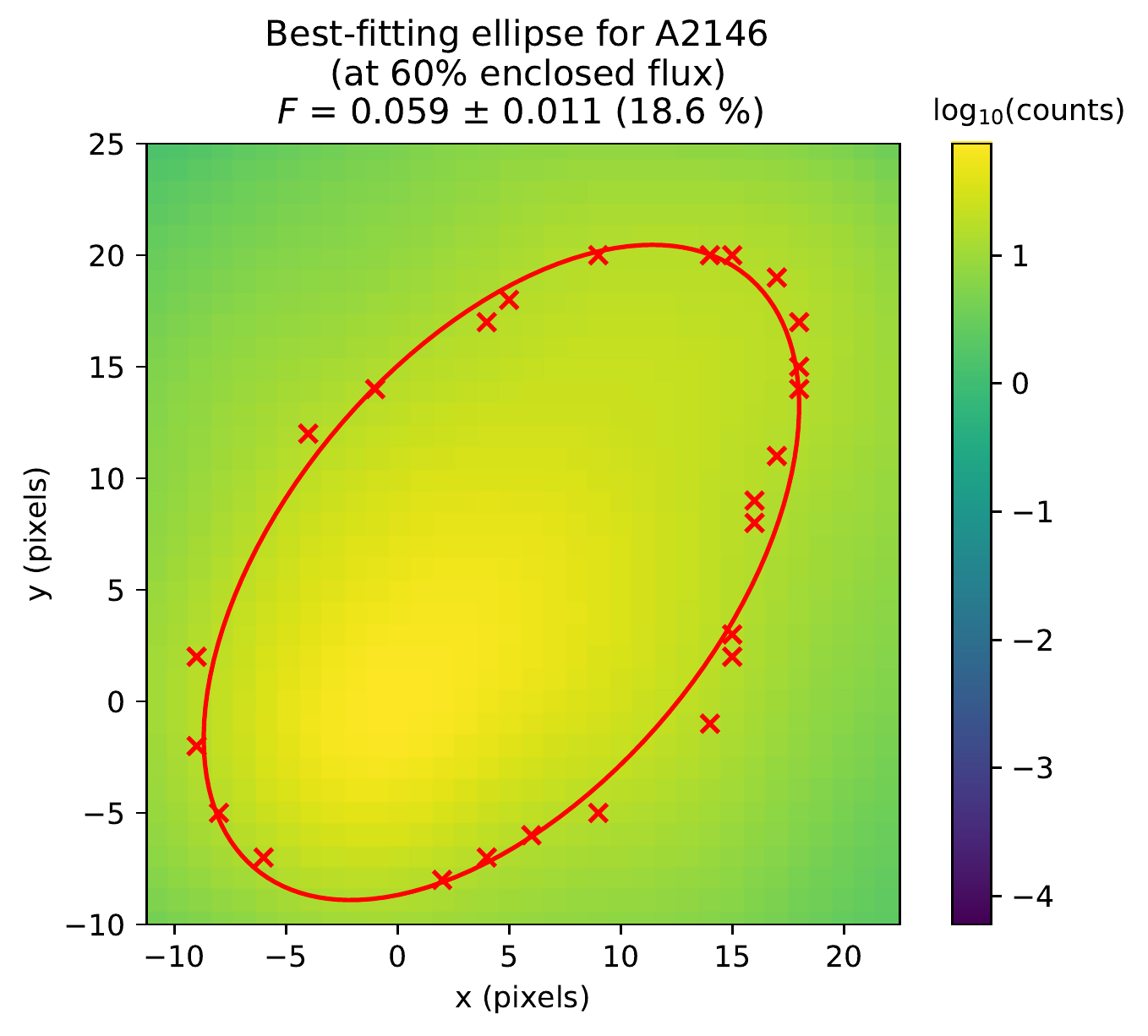}
\end{center} 
\caption{Smoothed X-ray surface brightness maps of \textit{Chandra}
  clusters A1201 and A2146, along with the isocontour points and
  best-fitting ellipse, for 60\% enclosed flux. These 2 clusters, among
  others, were not considered (rejected) for calculation of final PDF in
  this section, $\calP_{\rm obs}(F)$, because the error in filamentarity is greater than the bin width of $1/120 \sim 0.008 $ } 
\label{bad_clusters}
\end{figure}

\begin{figure}
\begin{center}
\includegraphics[width=\columnwidth]{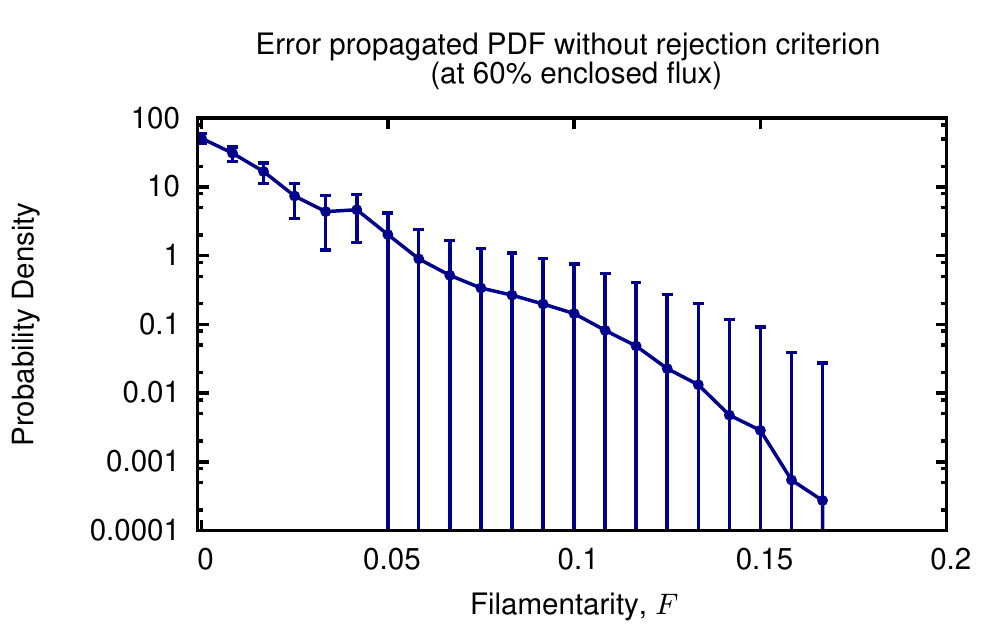}
\end{center} 
\caption{ PDF of filamentarity, $\calP_{\rm obs}(F)$, from
    \textit{Chandra} X-ray clusters for 60\% enclosed flux, obtained using 87
      clusters after propagation of isocontour errors, without the
      rejection criterion. The error bars show total error. The longer tail
      compared to Fig. \ref{error_propagated_PDF60} is due to few clusters
      with higher  isocontour noise, which leads to a spread in the
      filamentarity PDF. However, the filamentarity values in the
      additional bins in the long tail are all consistent with zero, while the non-zero values are the same as \ref{error_propagated_PDF60}.  }
\label{error_propagated_norejection}
\end{figure}

Thus, in addition to the Poisson error, we now have a new
component, the isocontour noise, in the final PDF $\calP_{\rm obs}(F)$. We
calculate the total error in each bin of the final PDF as: $\sigma_{total}^2 =
\sigma_{\rm Poisson}^2 + \sigma_{noise}^2$. The final PDF $\calP_{\rm
  obs}(F)$ for $60\%$ enclosed flux is shown in
Fig. \ref{error_propagated_PDF60}, which shows that the Poisson error gives the
dominant contribution in the total error.  The additional $\sigma_F<1/120$
selection criteria results in rejection of 4 clusters from our original
sample of 87 clusters for the $60\%$ flux case  in Sec.
\ref{chandra_analysis}. We show in Fig. \ref{bad_clusters} an example of 2
clusters which we have excluded using this criterion. We note that the
results would not be affected significantly even if we do not use this
additional rejection criterion and thus keep all the clusters for PDF
calculation. This is demonstrated for $\calP_{\rm obs}(F)$ of 60\%
enclosed flux in Fig. \ref{error_propagated_norejection}. The high $\sigma_F$ clusters would appear in different bins in
different realizations/sample of $P(F)$ resulting in spreading out of the
$P(F)$. In particular, the PDF  spreads out to higher values of
$F$ compared to Fig. \ref{error_propagated_PDF60}. However, all the higher filamentarity values have large error bars and
are consistent with zero, thereby keeping the final shape results
unchanged. The $P(F)$ values not consistent with zero remain the same and
are negligibly affected, whether we use the $\sigma_F$ rejection criterion or not.

This analysis method has two advantages: 
\begin{enumerate}
\item It allows us to obtain an estimate of  the error
  on final PDF due to the errors on ellipse axes lengths $(a,b)$ coming
  from the
  isocontour noise. The isocontour noise (or irregular shapes of
  isocontours)  maybe due to either noise in X-ray data (instrument noise
  or X-ray background) 
    or intrinsic irregularities in the cluster (departure from
    virialization or merging clusters).
\item It gives us an additional tool to clean our sample and thus reject,
  for example, merging clusters. However, merging clusters would anyway have large errorbars and therefore lower statistical weight as they
  will spread out their contribution over many bins. Thus these clusters
  automatically would be suppressed and as long as there are not too many
  of them, they will not affect the results.
\end{enumerate}

For our present analysis, however, we find that the Poisson errors dominate. In
the future however, for a larger sample of clusters, Poisson errors may
become sub-dominant. We have shown above, that in such a situation, we can
propagate errors from X-ray images in a robust manner. Our method, when
consistently propagating errors end-to-end, is particularly robust to
contamination of the sample by un-virialized and merging clusters.

\begin{figure}
\begin{center}
\includegraphics[scale=0.7]{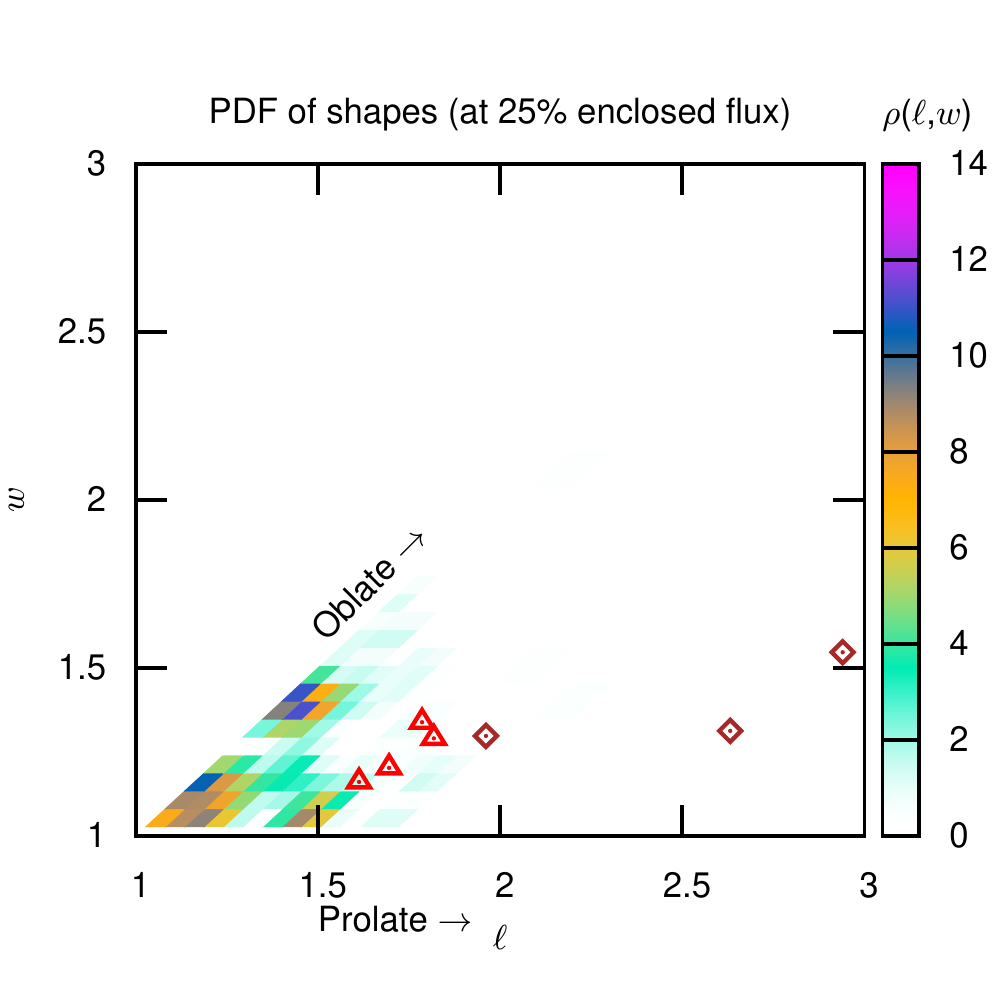}
\includegraphics[scale=0.7]{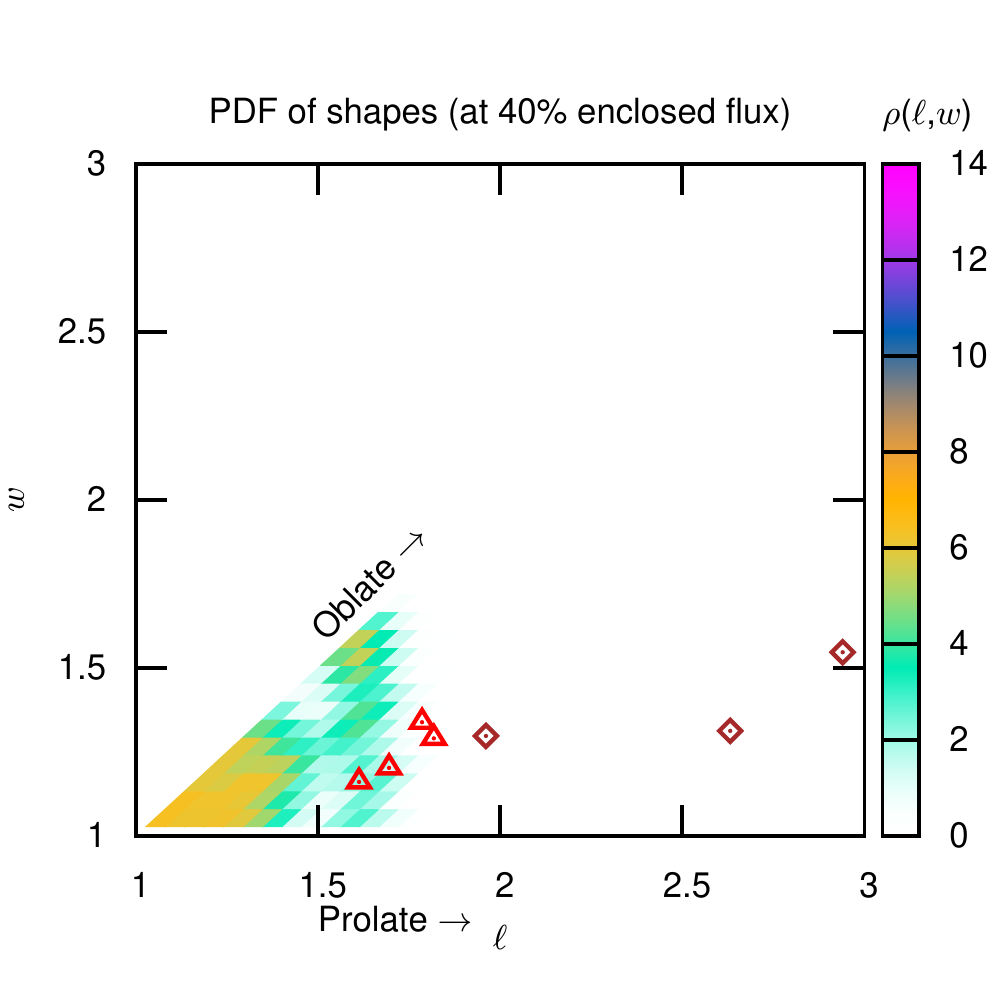}
\includegraphics[scale=0.7]{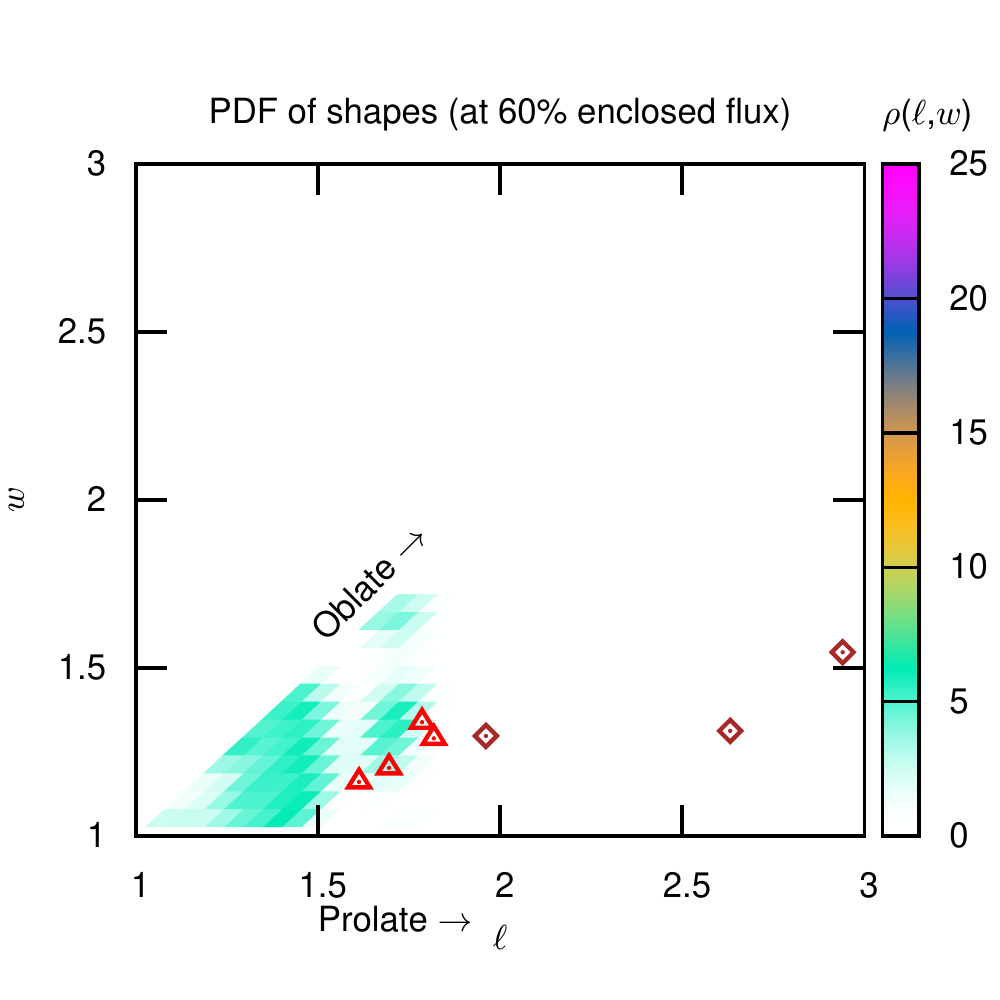}
\end{center} 
\caption{  2-D  probability density of shapes in the ($\ell,w$) plane
  obtained using the Monte Carlo method, same as
  Fig. \ref{shape_distribution}, but calculated using error propagated
  filamentarity PDF with rejection criterion. The shape PDF is close to the
  shape PDF in Fig. \ref{shape_distribution}, thus showing that inclusion
  of propagated isocontour errors and rejection criterion does not change the results significantly.} 
\label{shape_distribution_new}
\end{figure}

\begin{figure}
\begin{center}
\includegraphics[scale=0.7]{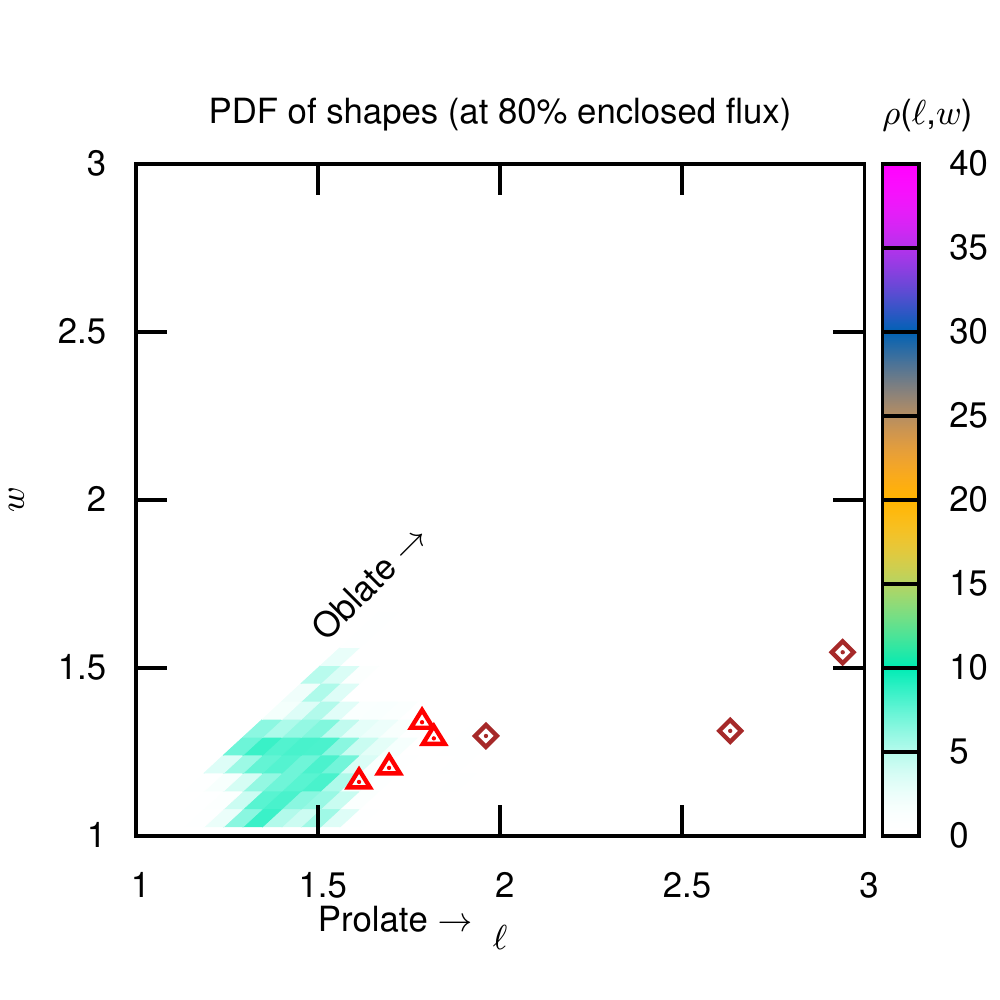}
\includegraphics[scale=0.7]{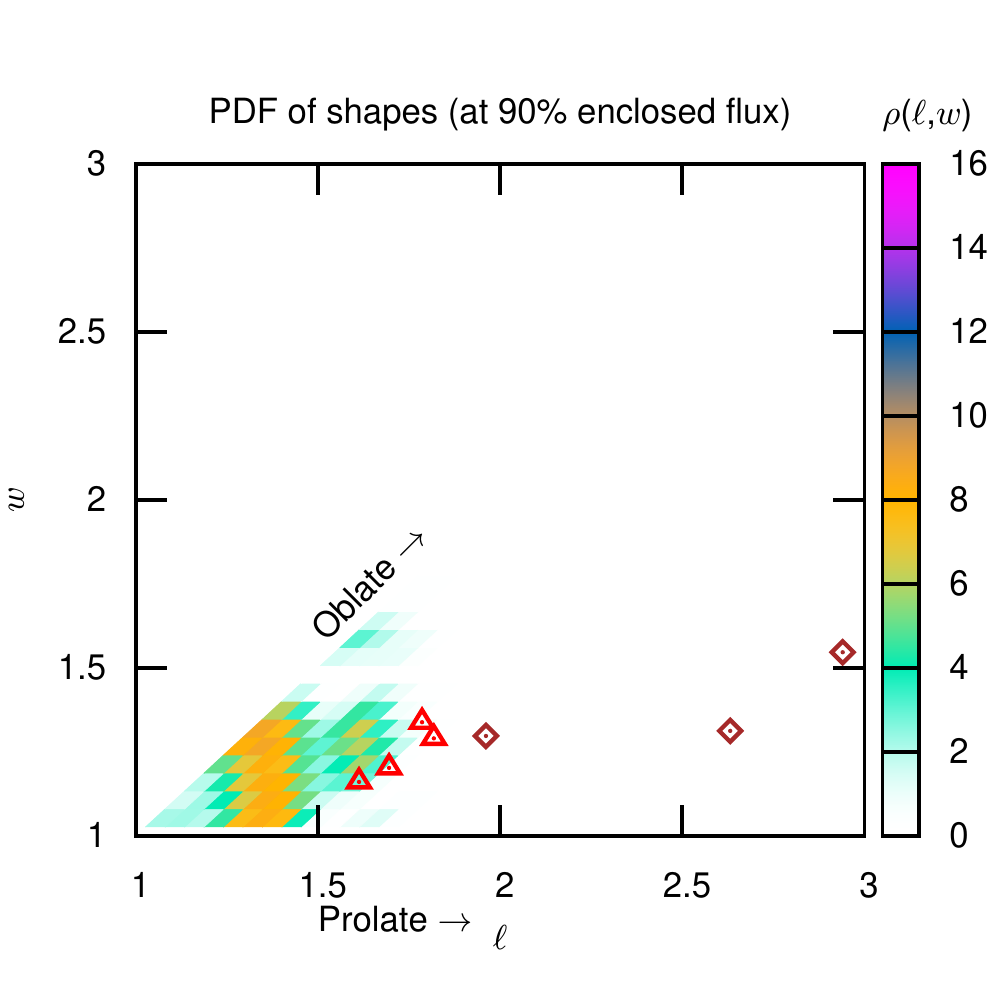}
\end{center} 
\caption{ Same as \ref{shape_distribution_new}, but for 80\% and 90\% enclosed flux } 
\label{shape_distribution_new2}
\end{figure}

We use the final error-propagated filamentarity PDF (including rejection
criterion) for Monte Carlo estimation of shape PDF, using the method
outlined in section \ref{Monte_carlo}. The rejection criterion leads to the rejection of 8, 6,
4, 4, and 11 clusters for 25\%, 40\%, 60\%, 80\%, and 90\% enclosed flux
respectively. The results from this analysis are
shown in Fig. \ref{shape_distribution_new} and
\ref{shape_distribution_new2}. We find that these shape PDFs are consistent
with the shape PDFs calculated in section \ref{Monte_carlo}. Thus, the results
do not change significantly after the inclusion of propagated errors from
isocontour noise  and additional rejection criterion, thereby showing the robustness of our method.

\newpage

\section{List of galaxy clusters and Chandra data used in this work}\label{clusterlist}
\begin{table}
\begin{minipage}{0.5\textwidth}
\centering
\caption{List of galaxy clusters and Chandra data used in this work.}
\label{chandralist}
\begin{tabular}{|c|c|c|}
\hline 
S.No. & Object name & Observation ID \\ 
\hline 
1 & A3532 & 10745 \\ 
\hline 
2 & 3C348 & 1625 \\ 
\hline 
3 & A3921 & 4973 \\ 
\hline 
4 & 4C55.16 & 1645 \\ 
\hline 
5 & cl1212+2733 & 5767 \\ 
\hline 
6 & cl0809+2811 & 5774 \\ 
\hline 
7 & cl0318-0302 & 5775 \\ 
\hline 
8 & cl0302-0423 & 5782 \\ 
\hline 
9 & 2PIGGz0.061 J0011.5-2850 & 5797 \\ 
\hline 
10 & 2PIGGz0.058 J2227.0-3041 & 5798 \\ 
\hline 
11 & A3102 & 6951 \\ 
\hline 
12 & cl1349+4918 & 9396 \\ 
\hline 
13 & AWM4 & 9423 \\ 
\hline 
14 & A0013 & 4945 \\ 
\hline 
15 & A0068 & 3250 \\ 
\hline 
16 & A0193 & 6931 \\ 
\hline 
17 & A0209 & 3579 \\ 
\hline 
18 & A0795  & 11734  \\ 
\hline 
19 & A0586 & 19962 \\ 
\hline 
20 & A0697 & 4217 \\ 
\hline 
21 & A0744  & 6947  \\ 
\hline 
22 &  A0773 &  5006 \\ 
\hline 
23 &  A0801 &  11767 \\ 
\hline 
24 &  A0907 & 3185  \\
\hline 
25 & A0963  &  903 \\
\hline 
26 & A1068  &  1652 \\
\hline 
27 &  A1201 &  4216 \\
\hline 
28 &  A1204 &  2205 \\
\hline 
29 &  A1361 &  2200 \\
 \hline 
30 & A1413  & 1661  \\
\hline 
31 & A1423  & 11724  \\
\hline 
32 & A1446  &  4975 \\
\hline 
33 & G125.70+53.85  & 15127  \\
\hline 
34 &  A1650 & 6356  \\
\hline 
35 &  A1664 &  7901 \\
\hline
\end{tabular}
\end{minipage}
\end{table}

\begin{table}
\begin{minipage}{0.5\textwidth}
\centering
\begin{tabular}{|c|c|c|}
\hline 
S.No. & Object name & Observation ID \\ 
\hline 
36 & A1763  &  3591 \\
\hline 
37 & A1835  & 495  \\
\hline 
38 &  A3827 & 7920  \\
\hline 
39 &  A1914 &  542 \\
\hline 
40 & A1930  &  A1930 \\
\hline 
41 & A2009  & 10438  \\
\hline 
42 &  A2029 &  891 \\
\hline 
43 & A2111  &  544 \\
\hline
44 &  A2124 &  3238 \\
\hline 
45 & A2142  &  5005  \\
\hline 
46 &  A2146  &  12246 \\
\hline 
47 & A2151  &  4996 \\
\hline 
48 & A2163  & 1653  \\
\hline 
49 &  A2187 &  9422 \\
\hline 
50 & A2199  &  10803 \\
\hline 
51 & A2204  & 499  \\
\hline 
52 & A2218  &  553 \\
\hline 
53 & A2219  &  14355 \\
\hline 
54 & A2244  &  4179 \\
\hline 
55 &  A2259 &  3245 \\
\hline 
56 &  A2261 & 550  \\
\hline 
57 &  A2276 & 10411  \\
\hline 
58 &  A2294 &  3246 \\
\hline 
59 & A2390  & 500  \\
\hline 
60 & A2409  &  3247 \\
\hline
61 & A2426  & 12279  \\
\hline 
62 &  A2445 &  12249 \\
\hline 
63 & A2485  &  10439 \\
\hline 
64 & A2537  &  4962 \\
\hline 
65 & A2550  &  2225 \\
\hline 
66 &  A2556 &  2226 \\
\hline 
67 &  A2589 &  6948 \\
\hline 
68 &  A2597 &  6934 \\
\hline 
69 &  A2626 &  3192 \\
\hline 
70 & A2631  &  3248 \\
\hline 
\end{tabular} 
\end{minipage}
\end{table}

\begin{table}
\centering
\begin{tabular}{|c|c|c|}
\hline 
S.No. & Object name & Observation ID \\ 
\hline 
71 &  A2717 &  6974 \\
\hline 
72 & A3112  &  2216 \\
\hline 
73 & A3528s   &  8268 \\
\hline 
74 & A3558  &  1646 \\
\hline 
75 & A4059  &  5785 \\
\hline 
76 & ACT J0616-5227  & 13127  \\
\hline 
77 &  AS1063 &  18611 \\
\hline 
78 & AWM7  &  908 \\
\hline 
79 & cl0956+4107  & 5759  \\
\hline 
80 & cl1120+4318  & 5771  \\
\hline 
81 & Cygnus A  & 360  \\
\hline 
82 &  ESO3060170-B & 3188  \\
\hline 
83 & MACSJ2311.5+0338  & 3288  \\
\hline 
84 &  PKS1404-257 &  1650 \\
\hline 
85 & SERSIC 159-03  &  1668 \\
\hline
86 & A383           &  2321 \\
\hline
87 & A1689          &  7289 \\
\hline
88 & MACS-J0329.6-0211  &  3582 \\
\hline
89 & RXJ1347.5-1145  &  3592 \\
\hline
\end{tabular}
\end{table}

\end{document}